\documentclass[letterpaper, 10 pt, conference]{ieeeconf}
\IEEEoverridecommandlockouts                              % This command is only
\usepackage{cite} % this converts [1], [2], [3] to [1]--[3]
 \usepackage[inkscapearea=page]{svg}
\usepackage{graphicx}      % include this line if your document contains figures
\usepackage{amsmath}
\usepackage{amssymb}
\usepackage{subcaption}

\usepackage{amsthm}
\usepackage{mathrsfs}
\usepackage{empheq}
\usepackage{tikz}
\usepackage{tikz-cd}
\usepackage[utf8]{inputenc}
\usepackage{pgfplots} 
\usepackage{pgfgantt}
\usepackage{pdflscape}
\pgfplotsset{compat=newest} 
\pgfplotsset{plot coordinates/math parser=false}
\usepackage{tikzscale}
\usepackage{standalone}
\usetikzlibrary{shapes,arrows}
\usetikzlibrary{arrows,calc,positioning,fit}
\usepackage{amsmath}
\let\labelindent\relax
\usepackage{enumitem}  % in the preamble

\usepackage{stfloats}% <-- added
\usepackage{hyperref}
\hypersetup{colorlinks = true,
 linkcolor = black,
 urlcolor = black,
 citecolor = black,
 anchorcolor = black}
 
\newtheorem{theorem}{Theorem}

\newtheorem{definition}{Definition}

\newtheorem{remark}{Remark}
\newtheorem{lemma}{Lemma}
\newtheorem{corollary}{Corollary}

\newtheorem{example}{Example}

\newtheorem*{question}{Question}

\def\BibTeX{{\rm B\kern-.05em{\sc i\kern-.025em b}\kern-.08em
    T\kern-.1667em\lower.7ex\hbox{E}\kern-.125emX}}
\begin{document}

\title{\LARGE \bf
On the IQC and SRG Frameworks: Equivalences and Converse Results
}
\author{Sebastiaan van den Eijnden, Chao Chen and Sei Zhen Khong% <-this % stops a space
\thanks{Sebastiaan van den Eijnden is with the Eindhoven University of Technology, Department of Mechanical Engineering, P.O. Box 513, 5600MB Eindhoven, The Netherlands (e-mail: s.j.a.m.v.d.eijnden@tue.nl).}
\thanks{Chao~Chen is with the Department of Electrical and Electronic Engineering, The University of Manchester, Manchester, M13 9PL, UK (e-mail: chao.chen@manchester.ac.uk).}
\thanks{Sei Zhen Khong 
is with the Department of Electrical Engineering, National Sun Yat-sen University, Kaohsiung 804201, Taiwan (e-mail: szkhong@mail.nsysu.edu.tw).}
}

\maketitle

\begin{abstract}
Integral quadratic constraints (IQCs) and scaled relative graphs (SRGs) are two prominent frameworks for analyzing feedback systems. Although both frameworks are rooted in the same graph separation principle, a general characterization of their relationship has remained unavailable. This paper establishes conditions under which the IQC- and SRG frameworks are equivalent. To this end, we introduce two new classes of systems, termed \emph{disjunctive} and \emph{conjunctive} systems, which are respectively characterized by unions and intersections of quadratic input–output constraints. Under structural assumptions, we prove equivalence between the IQC and SRG frameworks for these system classes. By exploiting this equivalence, we develop novel IQC- and SRG-based converse stability results for linear and nonlinear systems. Our results accommodate multiple quadratic constraints, extending well-known converse results in robust control beyond the single-constraint setting.    
\end{abstract}

\section{Introduction}

The analysis of feedback systems is often grounded in methods for predicting the qualitative behaviour of a feedback interconnection using properties of its individual components. Loosely speaking, a feedback interconnection is stable provided the input-output graphs of the interconnected components are suitably separated \cite{Teel96}. A variety of useful sufficient conditions for graph separation, and hence stability, have been developed based on \emph{quadratic} separation techniques. To be specific, the open-loop system graphs are characterized by complementary quadratic forms, from which closed-loop stability follows. Classical examples include small-gain, passivity, and conicity theorems \cite{Zames66, Willems71, Desoer09}, whereas more general and recent results follow from integral quadratic constraints (IQCs) \cite{Megretski, Khong21IQC, Su2} and scaled relative graphs (SRGs) \cite{Chaffey23, Eijnden24, Chen26}.  

Both the IQC and SRG frameworks provide realizations of the quadratic graph separation principle, but they emphasize different perspectives. The close connection between IQCs and linear matrix inequalities (LMIs) via the Kalman--Yakubovich--Popov lemma \cite{Rantzer} and Iwasaki--Hara lemma~\cite{IwasakiHara05} has enabled many powerful computational tools for the analysis and design of robust control systems; see, e.g., \cite{Veenman14, Veenman16, Scherer22, KhongLanzon24, Schwenkel26}. In contrast, SRGs in a complex plane emphasize a geometric and graphical viewpoint based on gain-phase relationships, akin to the Nyquist diagram for linear time-invariant (LTI) systems. They offer intuitive insights into feedback interconnections and their robustness properties \cite{Chaffey23, Pates, Eijnden24, Chen26, Krebbekx25, Krebbekx25a}. 

Despite their common foundation in quadratic graph separation, the relationship between IQCs and SRGs is yet not fully understood. Initial connections have been established in \cite{Chaffey23}, \cite{Groot25}, \cite{Groot26} and \cite{Cha25}, but a general characterization linking the two frameworks is still lacking. Developing such a link is important for nonlinear systems, as it brings together the computational power of IQCs and the graphical intuition of SRGs. 

In this paper, we establish conditions under which the IQC and SRG frameworks are equivalent. In doing so, within both frameworks we introduce two novel and fundamentally different classes of systems: \emph{disjunctive} systems and \emph{conjunctive} systems. Disjunctive systems are characterized by \emph{unions} of quadratic input-output constraints, whereas conjunctive systems are characterized by \emph{intersections} of such constraints. The terminology is borrowed from classical logic, where disjunctive and conjunctive correspond to logical OR and logical AND, respectively. Here, these terms reflect the underlying union- and intersection-based system characterizations, which arise naturally in many relevant settings such as gain-scheduled, switched, and mixed systems \cite{Leith, Eijnden24, Griggs, EijCha, Eij26}. We show that for the new system classes, under structural assumptions, the IQC and SRG graph separation conditions are equivalent, as summarized below by Fig.~\ref{fig:equivalence}. We initially focus on soft and static IQCs and SRGs, and comment on extensions to their hard counterparts and dynamic multipliers. 

An immediate consequence of the established equivalence is that any result obtained in one framework can be carried over directly to the other. This is particularly useful in the derivation of converse results for feedback stability analysis. Various IQC type converse results have been established recently in \cite{Khong18, Khong21, Khong21b, Khong24, Liu25, Ringh26, Khong26, Yang, Zhang26}, and play an important role in robustness analysis by providing necessary and sufficient (non-conservative) conditions for robust stability under specific uncertainty classes. Interestingly, all of the above results pertain to systems involving a single property or quadratic type of constraint. By leveraging the disjunctive and conjunctive system classes, we establish, for the first time, converse SRG theorems for nonlinear feedback stability analysis. Through the established IQC–SRG equivalence, we translate these converse results into the IQC framework, yielding new converse results for LTI and nonlinear systems involving multiple constraints. Our results extend the scope of non-conservative robustness analysis. The proofs are constructive and reveal a structural challenge in establishing converse theorems for conjunctive system classes, i.e., when dealing with intersections of constraints. 

The remainder of the paper is organized as follows. Notation and mathematical preliminaries are presented in Section~\ref{sec:preliminaries}. Section~\ref{sec:SRG_IQC} reviews relevant background on IQCs and SRGs. Our first main result is presented in Section~\ref{sec:equivalence}, where we establish conditions under which the IQC and SRG frameworks are equivalent. We present our second main result in Section~\ref{sec:converse} in the form of converse IQC and SRG stability theorems. Section~\ref{sec:conclusions} concludes the paper.  

\section{Notation and Preliminaries}\label{sec:preliminaries}

Let $\mathbb{R}$ and $\mathbb{C}$ denote the set of real and complex numbers, and let $\mathbb{R}^{m\times n}$ and $\mathbb{C}^{m \times n}$ denote the set of real and complex matrices of size $m \times n$, respectively. We define the open unit disk in $\mathbb{C}$ by $\mathbb{D}:=\left\{z \in \mathbb{C} \mid |z|<1\right\}$. For $c\in\mathbb{R}$ and $r>0$, we denote by $B(c,r):=\{z\in\mathbb{C}\mid |z-c|<r\}$ the open ball in $\mathbb{C}$ centered at $c$ with radius $r$. The distance between two sets ${S}_1,S_2\subseteq \mathbb{C}$ is denoted by $\textup{dist}({S}_1,{S}_2)=\inf_{a\in {S}_1,\;b\in{S}_2}|a-b|$. The cardinality of a set $S$, i.e., the number of its elements, is denoted by $|S|$. The closure of a set $S \subseteq \mathbb{C}$ is denoted as $\overline{S}$. The superscript $^\top$ is used to denote transposition.

\subsection{Signal Spaces}
Let $\mathcal{F}^n$ denote the space of all functions mapping the interval $[0,\infty)$ into $\mathbb{R}^n$, where the superscript $n$ is dropped whenever the dimension is clear from the context. Denote the space of Lebesgue square-integrable functions $u\in \mathcal{F}^n$ by
$$\mathcal{L}_2^n := \left\{u\in \mathcal{F}^n \;\middle|\; \int_{0}^\infty u(t)^\top u(t) dt < \infty\right\}.$$ 
For signals $u,y \in \mathcal{L}_{2}^n$, let the inner product and norm be {defined by}
$$\langle u,y\rangle := \int_{0}^\infty u(t)^\top y(t) dt \:\:\textup{ and }\:\:\|u\| := \sqrt{\langle u,u\rangle},$$
respectively. Given $u \in \mathcal{L}_2^n$, let $\hat{u}$ denote its Fourier transform, that is,
$$\hat{u}(j\omega):=\int_{0}^{\infty}u(t)e^{-j\omega t}\,dt, \quad \omega \in \mathbb{R}.$$

The space of square-integrable functions on finite time intervals is defined by 
\begin{equation*}
     \mathcal{L}_{2e}^n :=\left\{u\in \mathcal{F}^n \;\middle|\; \int_{0}^T u(t)^\top u(t) dt < \infty, \: \forall T \geq 0\right\},
\end{equation*}
which is also referred to as the extended $\mathcal{L}_2$-space. 

\subsection{Systems}
In this paper, a \emph{system} is specified by a single-valued input-output map $H:\mathcal{L}_{2e}^m \to \mathcal{L}_{2e}^n$, satisfying $H(0)=0$. The $\mathcal{L}_2$-domain of $H$ is denoted by $\textup{dom}(H):=\left\{u \in \mathcal{L}_2^m \;\middle|\; H(u) \in \mathcal{L}_2^n\right\}$, i.e., the set of all signals in $\mathcal{L}_2^m$ that are mapped to signals in $\mathcal{L}_2^n$. The \emph{graph} and \emph{inverse graph} of a system $H$ are, respectively, given by
\begin{equation*}
  \mathcal{G}(H):= \left\{\begin{bmatrix}
        u \\ H(u)
    \end{bmatrix}\in \mathcal{L}_2^m \times \mathcal{L}_2^n\right\}, 
\end{equation*}
and 
\begin{equation*}
  \mathcal{G}^{-1}(H):= \left\{\begin{bmatrix}
        H(u) \\ u
    \end{bmatrix}\in \mathcal{L}_2^n \times \mathcal{L}_2^m\right\}. 
\end{equation*}

The notation $H_1(H_2)$ and $H_1+H_2$ is used to denote the cascade and sum of two operators $H_1$ and $H_2$, i.e., $y = H_1(H_2(u))$ and $y = (H_1+ H_2)(u) = H_1(u) + H_2(u)$.

Define the \emph{truncation} operator $\Gamma_T:\mathcal{F} \to \mathcal{F}$ to be $(\Gamma_T(u))(t)=u(t)$ when $t\leq T$, and $(\Gamma_T(u))(t)=0$ when $t> T$. We say that $H:\mathcal{L}_{2e}^m \to \mathcal{L}_{2e}^n$ is \emph{causal} if $\Gamma_T H \Gamma_T = \Gamma_T H$ for all $T\geq 0$. Likewise, define the \emph{right shift} operator $S_T:\mathcal{F} \to \mathcal{F}$ to be $(S_T (u))(t) = u(t-T)$ for $t\geq T$ and $(S_T (u))(t) = 0$ for $0\leq t<T$. We say that $H:\mathcal{L}_{2e}^m \to \mathcal{L}_{2e}^n$ is \emph{time-invariant} if $S_T H = HS_T$ for all $T \geq 0$. Finally, we say that $H$ is \emph{linear} if $H(a_1 u_1+a_2 u_2) = a_1H(u_1)+a_2 H(u_2)$ for all $u_1,u_2 \in \mathcal{L}_{2e}^m$ and $a_1, a_2 \in \mathbb{R}$. For a linear and time-invariant system $H$, we will write with some abuse of notation $H(s)$, $s \in \mathbb{C}$ and $H(j\omega)$, $\omega \in \mathbb{R}$ to denote its transfer function and frequency-response function, respectively.

Throughout this paper, we consider ``square'' systems, i.e., systems having the same number of inputs and outputs ($m=n$). This is done without loss of generality, as a system may always be patched with zero input/output channels to make it ``square''. The following definitions of stability and (incremental) boundedness are adopted.  

\begin{definition}
     A causal system $H:\mathcal{L}_{2e}^n \to \mathcal{L}_{2e}^n$ is said to be 
    \begin{itemize}
        \item \emph{stable} if $\textup{dom}(H)=\mathcal{L}_2^n$;
        \item \emph{bounded} if it is stable and \begin{equation*}
        \|H\|:=\sup_{0\neq u\in \mathcal{L}_2^n} \frac{\|H(u)\|}{\|u\|}<\infty;
    \end{equation*}
    \item \emph{incrementally bounded} if it is stable and
    \begin{equation*}
        \|H\|_\Delta := \sup_{u,v\in \mathcal{L}_2^n, u\neq v}\frac{\|H(u)-H(v)\|}{\|u-v\|}<\infty.
    \end{equation*}
    \end{itemize}
    We refer to $\|H\|$ as the $\mathcal{L}_2$-gain and $\|H\|_\Delta$ as the incremental $\mathcal{L}_2$-gain of $H$. 
\end{definition}

The sets of bounded and incrementally bounded causal systems are denoted by $\mathcal{B}$ and $\mathcal{B}_\Delta$, respectively. When $H$ is linear, it is well-known that $\|H\|=\|H\|_\Delta$. Let $\mathcal{L}$ denote the set of causal and bounded LTI systems. Clearly, $\mathcal{L}\subseteq\mathcal{B}_\Delta \subseteq\mathcal{B}$. Moreover, an LTI system $H$ is causal and bounded if its transfer function matrix $H(s)$ is real, rational, stable, and proper, where stability in this context means that all poles lie in the open left-half complex plane and properness means that $H(\infty)$ is finite. The space of such $n\times n$ transfer function matrices is denoted by $\mathcal{RH}_{\infty}^{n\times n}$.  

\subsection{Feedback Systems and Stability}
\tikzset{
    block/.style = {draw, rectangle,
        minimum height=1cm,
        minimum width=1cm},
    input/.style = {coordinate,node distance=1cm},
    output/.style = {coordinate,node distance=1cm},
    arrow/.style={draw, -latex,node distance=2cm},
    sum/.style = {draw, circle, node distance=.5cm, inner sep=.3em},
}      

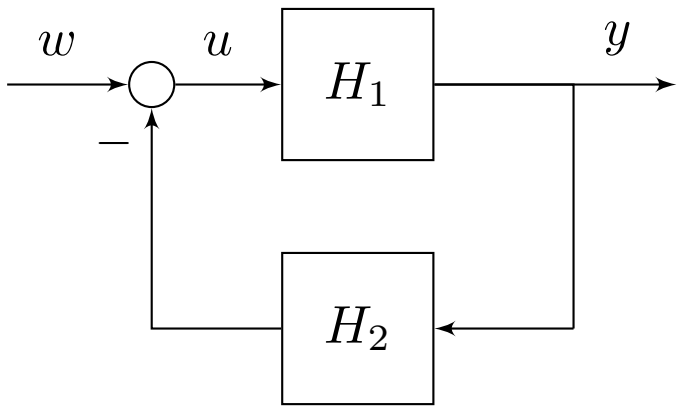
\begin{figure}[tb!]
\centering
\begin{tikzpicture}[auto, node distance=1cm,>=latex']
\centering
\node [block] (H1) {$H_1$};
\node [sum, left=0.7 of H1] (s1) {};         
\node [block, below=.6 of H1] (H2) {$H_2$};
%\node [sum, right=.7 of H2] (s2) {}; 
\coordinate[right=.92 of H2] (s2);
 
\draw[->] (s1) --node [xshift = -2, yshift = 2 , pos=0.5] {$u$}  (H1);
\draw[->] (s2) --node [xshift = 2, yshift = -2 , pos=0.5] {}  (H2);

\draw[-] (H1) -| node [pos=0.92] { } (s2);

\draw[->] (H2) -| node [pos=0.92] {\small $-$} (s1);

\node [input, left=.8 of s1] (dummy1){};
\draw[->] (dummy1) --node [xshift = -2, yshift = 2 , pos=0.5] {$w$}(s1);

\node [input, right=1.6 of H1] (dummy4){};
\draw[->] (H1) --node [xshift = -2, yshift = 2 , pos=0.8] {$y$}(dummy4);

\end{tikzpicture}
 \caption{Feedback interconnection.}
 \label{fig:FB1}
\end{figure}
We are primarily interested in the feedback interconnection of two causal systems $H_1 : \mathcal{L}_{2e}^n \to \mathcal{L}_{2e}^n$ and $H_2:\mathcal{L}_{2e}^n \to \mathcal{L}_{2e}^n$, as depicted in Fig.~\ref{fig:FB1} and described by the functional equations:
\begin{equation}\label{eq:system}
\begin{aligned}
    u &= w - H_2(y),\\
    y &= H_1( u).
\end{aligned}
\end{equation}
Let $[H_1, H_2]$ denote this feedback system, and let the mapping from $u$ to $w$ be denoted by
\begin{equation}\label{eq:Mapping}
 F^\dagger_{[H_1, H_2]}:=I+H_2(H_1):\mathcal{L}_{2e}^{n} \to \mathcal{L}_{2e}^{n}.
\end{equation}

In the remainder, we assume that the feedback system in \eqref{eq:system} is \emph{well-posed} in the sense that $F^\dagger_{[H_1, H_2]}$ in \eqref{eq:Mapping} has a causal inverse on $\mathcal{L}_{2e}^{n}$. This assumption allows us to write the mapping from $w$ to $u$ on $\mathcal{L}_{2e}^{n}$ as
\begin{equation}
   F_{[H_1,H_2]}:= (F^\dagger_{[H_1,H_2]})^{-1}= \left( I+H_2(H_1)\right)^{-1}.
\end{equation}
Concluding on stability and (incremental) boundedness of the feedback system $[H_1, H_2]$ then amounts to showing that 
\begin{itemize}
    \item [(a)] $\textup{dom}(F_{[H_1,H_2]}) = \mathcal{L}_{2}^{n}$;
    \item [(b)] $\|F_{[H_1,H_2]}\|<\infty$ (respectively $\|F_{[H_1,H_2]}\|_{\Delta}<\infty$).
\end{itemize}
 The next result from \cite{Cha25} presents conditions for incremental boundedness. Analogous results can be formulated for the non-incremental case. 

\begin{theorem}[\hspace{1sp}{\cite[Theorem 2]{Cha25}}]\label{th:inc}
Consider a feedback interconnection $[H_1,H_2]$ in \eqref{eq:system}. Suppose the following holds: 
    \begin{enumerate}[label=(\roman*)]     \renewcommand{\theenumi}{{(\roman{enumi})}}\renewcommand{\labelenumi}{\theenumi}
        \item $H_1, H_2 \in \mathcal{B}_\Delta$; \label{item_Thm1_boundedness}
        \item $[H_1, \tau H_2]$ is well-posed for all $\tau \in (0,1]$; \label{item_Thm2_wellposedness}
        \item there exists a finite number $\gamma >0$ such that, for all $\tau \in [0,1]$ and all $w_1, w_2 \in \textup{dom}(F_{[H_1, \tau H_2]})$, we have
        \begin{equation*}
            \|u_1-u_2\| \leq \gamma \|w_1-w_2\|,
        \end{equation*}
        where $u_i = F_{[H_1, \tau H_2]}(w_i)$, $i=1,2$.  \label{item_Thm2_separation}
    \end{enumerate}
    Then, the feedback interconnection $[H_1,H_2]$ is incrementally bounded with $\|F_{[H_1, H_2]}\|_\Delta = \gamma$.
\end{theorem}

Items~\ref{item_Thm1_boundedness} and \ref{item_Thm2_wellposedness} of Theorem~\ref{th:inc} are typically satisfied by inherent physical properties. By contrast, item \ref{item_Thm2_separation} can be verified using several operator theoretic tools. We consider two closely related tools to achieve this: IQCs and SRGs. 

\section{IQCs and SRGs}\label{sec:SRG_IQC}
In this section, we introduce the notions of IQCs and SRGs and provide some basic results on feedback systems.

\subsection{Integral Quadratic Constraints (IQCs)}
Incremental IQCs \cite{Megretski} provide a way to constrain incremental input--output relationships of systems. Let $\Delta(\cdot)$ denote
the difference between two signals labeled by $(\cdot)_1$ and $(\cdot)_2$, e.g., $\Delta x = x_1-x_2$.  
 
\begin{definition}[Incremental IQC]\label{def:IQC}
  Consider a causal and stable system $H:\mathcal{L}_{2e}^n \to \mathcal{L}_{2e}^n$. Let $M \in \mathbb{R}^{2n\times 2n}$ be symmetric. Then, $H$ is said to satisfy the \emph{incremental IQC defined by the multiplier $M$} if 
  \begin{equation}\label{eq:sIQC}
    \sigma_\Delta(H,M):=  \int_{0}^\infty \begin{bmatrix}
          \Delta u(t)\\\Delta y(t)
      \end{bmatrix}^\top M \begin{bmatrix}
          \Delta u(t)\\\Delta y(t)
      \end{bmatrix} dt \geq 0
  \end{equation}
  for all $(u_1,y_1), (u_2,y_2) \in \mathcal{G}(H)$.
\end{definition}

Let $\mathbf{M}$ be a collection of a \emph{finite} number of distinct symmetric matrices $M\in \mathbb{R}^{2n\times 2n}$, i.e., $|\mathbf{M}| = l$ and 
\begin{equation}\label{eq:set_multiplier_M}
\mathbf{M} := \{M_k \in \mathbb{R}^{2n\times 2n}\mid M_k=M_k^\top, k = 1, 2, \ldots,  l \}.  
\end{equation}
Define the following two system classes based on IQC characterizations:
\begin{equation}\label{eq:HI}
\begin{split}
    {\mathcal{H}}^\cup_{\textup{IQC}}(\mathbf{M})\!:=\!\big\{&H\in \mathcal{B}_\Delta \mid \:\forall (u_1,y_1), (u_2,y_2) \in \mathcal{G}(H),\\
    &\qquad\quad \exists M \in \mathbf{M} \:\:\textup{s.t.} \:\sigma_\Delta(H,M)\geq 0\big\},
    \end{split}
\end{equation}
and
\begin{equation}\label{eq:HI2}
\begin{split}
    {\mathcal{H}}^\cap_{\textup{IQC}}(\mathbf{M})\!:=\!\big\{&H\in \mathcal{B}_\Delta \mid \:\forall (u_1,y_1), (u_2,y_2) \in \mathcal{G}(H),\\
    &\quad\qquad\quad \forall M \in \mathbf{M}, \:\sigma_\Delta(H,M)\geq 0\big\}.
    \end{split}
\end{equation}

The two system classes $\mathcal{H}^\cup_{\mathrm{IQC}}(\mathbf{M})$ and
$\mathcal{H}^\cap_{\mathrm{IQC}}(\mathbf{M})$ capture two fundamentally \emph{different} characterizations with respect to the multiplier family $\mathbf{M}$.
The class $\mathcal{H}^\cup_{\mathrm{IQC}}(\mathbf{M})$ describes a
\emph{disjunctive (union-type)} property: a system in this class may satisfy \emph{at least
one} IQC from the given family for each input-output trajectory. In contrast, the class $\mathcal{H}^\cap_{\mathrm{IQC}}(\mathbf{M})$
represents a \emph{conjunctive (intersection-type)} property, where \emph{all} IQCs in the family
are simultaneously satisfied for all input-output trajectories. Clearly, it follows that $$\mathcal{H}^\cap_{\textup{IQC}}(\mathbf{M})\subseteq\mathcal{H}^\cup_{\textup{IQC}}(\mathbf{M}).$$ 

The next results provide IQC-based tests for verifying item~\ref{item_Thm2_separation} in Theorem~\ref{th:inc} when one of the feedback components in $[H_1, H_2]$ belongs to either the class $\mathcal{H}^\cup_{\textup{IQC}}(\mathbf{M})$ or $\mathcal{H}^\cap_{\textup{IQC}}(\mathbf{M})$. 

{
\renewcommand{\thetheorem}{\number\numexpr\value{theorem}\relax.A}
\begin{theorem}\label{lem:02}
Given $\mathbf{M}$ as in \eqref{eq:set_multiplier_M}, and consider a feedback interconnection $[H_1,H_2]$. Suppose the following holds:  
 \begin{enumerate}[label=(I\arabic*.A),leftmargin=*]
        \item $H_1\in \mathcal{B}_\Delta$ and $\tau H_2 \in \mathcal{H}^\cup_{\textup{IQC}}(\mathbf{M})$ for all $\tau \in (0,1]$; \label{item_Thm2_I1A}
        \item $[H_1, \tau H_2]$ is well-posed for all $\tau \in (0,1]$; \label{item_Thm2_I2A}
        \item for all $M \in \mathbf{M}$, there exists $\epsilon >0$ such that 
    \begin{equation}\label{eq:IQC_FDt0}
      \int_{0}^\infty  \begin{bmatrix}
          \Delta y_1(t)\\ \Delta u_1(t)
        \end{bmatrix}^\top (M+\epsilon I) \begin{bmatrix}
            \Delta y_1(t)\\ \Delta u_1(t)
        \end{bmatrix}dt \leq 0, 
    \end{equation}
   holds for all $(y_1,u_1), (y_2,u_2)\in \mathcal{G}^{-1}(-H_1)$.  \label{item_Thm2_I3A}
    \end{enumerate}
    Then, the interconnection $[H_1,H_2]$ is incrementally bounded.
\end{theorem}
}

\begin{proof}
Let $(\Delta u^{(1)},\Delta y^{(1)})$ denote the incremental input-output pair corresponding to $H_1$, and let $(\Delta u^{(2)},\Delta y^{(2)})$ denote the incremental input-output pair for $\tau H_2$. Note from the system interconnection in \eqref{eq:system} that $\Delta u^{(1)} = \Delta w - \Delta y^{(2)}$, and $\Delta y^{(1)} = \Delta u^{(2)}$. From supposition $(\textup{I}1.\textup{A})$ it follows that for each input-output pair $(\Delta u^{(2)}, \Delta y^{(2)})$ there is an $M \in \mathbf{M}$ such that
\begin{equation}\label{eq:pf10}
    \int_{0}^\infty \begin{bmatrix}
          \Delta u^{(2)}(t)\\\Delta y^{(2)}(t)
      \end{bmatrix}^\top M \begin{bmatrix}
          \Delta u^{(2)}(t)\\\Delta y^{(2)}(t)
      \end{bmatrix} dt \geq 0.
\end{equation}
By adding \eqref{eq:IQC_FDt0} to \eqref{eq:pf10}, and using the system interconnection as well as boundedness of the elements of $M$, it can be verified from, e.g., \cite[Lemma 7]{Cha25} that for the corresponding $w_1,w_2 \in \textup{dom}(F_{[H_1,\tau H_2]})$ there exists $c_w>0$, independent of $\tau$, such that
$$
\|\Delta u\| \leq c_w \|\Delta w\|.
$$
Repeating the above steps to cover all $w_1,w_2 \in \textup{dom}(F_{[H_1,\tau H_2]})$, and taking $\gamma$ as the supremum over all the corresponding $c_w$, we establish item $\ref{item_Thm2_separation}$ in Theorem~\ref{th:inc}. Since $(\textup{I}1.\textup{A})$ and $(\textup{I}2.\textup{A})$ establish the remaining items in Theorem~\ref{th:inc}, the result follows. 
\end{proof} 

{
\renewcommand{\thetheorem}{\number\numexpr\value{theorem}\relax.B}
\addtocounter{theorem}{-1} % keep counter from advancing
\begin{theorem}\label{lem:2}
Given $\mathbf{M}$ as in \eqref{eq:set_multiplier_M}, and consider a feedback interconnection $[H_1,H_2]$. Suppose the following holds:
 \begin{enumerate}[label=(I\arabic*.B),leftmargin=*]
        \item $H_1\in \mathcal{B}_\Delta$ and $\tau H_2 \in \mathcal{H}^\cap_{\textup{IQC}}(\mathbf{M})$ for all $\tau \in (0,1]$; \label{item_Thm2_I1B}
        \item $[H_1, \tau H_2]$ is well-posed for all $\tau \in (0,1]$; \label{item_Thm2_I2B}
        \item there exist $\mu_{M} \geq 0$, not all zero, and $\epsilon >0$, such that
    \begin{equation}\label{eq:IQC_FD}
      \int_{0}^\infty  \begin{bmatrix}
          \Delta y_1(t)\\ \Delta u_1(t)
        \end{bmatrix}^\top \!\!\!\!\Big(\sum_{M \in \mathbf{M}}\mu_MM+\epsilon I\Big) \begin{bmatrix}
            \Delta y_1(t)\\ \Delta u_1(t)
        \end{bmatrix}dt \leq0, 
    \end{equation}
    for all $(y_1,u_1), (y_2,u_2)\in \mathcal{G}^{-1}(-H_1)$. \label{item_Thm2_I3B}
    \end{enumerate}
    Then, the interconnection $[H_1,H_2]$ is incrementally bounded.
\end{theorem}
}

\begin{proof}
Construct the multiplier $\tilde{M}=\sum_{M\in\mathbf{M}}\mu_M M$ with $\mu_M\geq 0$. Since $\tilde{M}$ defines a valid incremental IQC for $\tau H_2$, the conclusion follows directly from \cite[Corollary 2]{Cha25}.
\end{proof}

The significance of making a distinction between both the disjunctive and conjunctive system classes will become clear later in Section~\ref{sec:equivalence} and Section~\ref{sec:converse} where we establish equivalence results and converse theorems. 

\subsection{Scaled Relative Graphs (SRGs)}\label{sub:SRG}
SRGs \cite{Chaffey23} offer a graphical representation of the input--output behaviour of a system in terms of incremental gain and incremental phase. Let $u,y\in \mathcal{L}_2^n$. Define the gain $\rho(u,y)\in [0,\infty]$ from $u$ to $y$ by
\begin{equation}\label{eq:gain}
    \rho(u,y):=\frac{\|y\|}{\|u\|} 
\end{equation}
if $u\neq 0$, and $\rho(u,y):=\infty$ otherwise, and the phase $\theta(u,y)\in [0,\pi]$ from $u$ to $y$ by
\begin{equation}\label{eq:phase}
    \theta(u,y):=\arccos \frac{\langle u,y\rangle}{\|u\|\|y\|}
\end{equation}
if $u,y\neq 0$, and $\theta(u,y):=0$ otherwise. 

For $u,y\in\mathcal{L}_2^n$ with $u\neq 0$, we write $(u,y)\mapsto z$ to indicate the complex numbers $z(u,y) = \rho(u,y)e^{\pm j\theta(u,y)}$ generated by the signal pair $u,y$. For a system $H$, the collection of all complex numbers generated by incremental input-output pairs makes up the SRG.

\begin{definition}[SRG]\label{eq:SG}
    Consider a causal and stable system $H:\mathcal{L}_{2e}^n \to \mathcal{L}_{2e}^n$. The \emph{SRG} of $H$ is defined as
\begin{equation}\label{eq:sSRG}
\begin{split}
    &\textup{SRG}(H):=\big\{\rho(\Delta u,\Delta y)e^{\pm j\theta(\Delta u,\Delta y)}\in \mathbb{C}\mid\\
    & \qquad  (u_1,y_1),(u_2,y_2)\in \mathcal{G}(H), \;\Delta u \neq 0,~\Delta y \neq 0\big\}.
    \end{split}
\end{equation}
\end{definition}

In the remainder, we will also work with the \emph{inverse SRG} of $H$, which is defined as
\begin{equation}\label{eq:SGi}
    \textup{SRG}(H)^{-1}:=\left\{z^{-1} \in \mathbb{C} \;\middle|\; z \in \textup{SRG}(H)\right\}.
\end{equation}
Notice that $\textup{SRG}(H)$ is always symmetric about the real axis. We now characterize a region in the complex plane that contains $\textup{SRG}(H)$. Let 
\begin{equation}\label{eq:set_complex_region}
\mathcal{S} \subseteq \mathbb{C},\;\;\mathcal{S}=\{z^* \:\mid \: z\in \mathcal{S}\},
\end{equation}
i.e., the set $\mathcal{S}$ is symmetric about the real axis. Similar as for IQCs, given a set $\mathcal{S}$, we define a system class via SRG characterizations:
\begin{equation}\label{eq:HS}
\begin{split}
    \mathcal{H}_{\textup{SRG}}(\mathcal{S}):=\big\{&H\in \mathcal{B}_\Delta \:\mid \:\textup{SRG}(H) \subseteq \mathcal{S}\big\}.
    \end{split}
\end{equation}
Note that no convexity assumptions are imposed on the set $\mathcal{S}$; it may be convex or non-convex and may arise through unions and intersections of other sets.

The next result parallels Theorems~\ref{lem:02} and \ref{lem:2} and provides an SRG-based test for verifying item \ref{item_Thm2_separation} in Theorem~\ref{th:inc}, when one of the feedback components in $[H_1,H_2]$ belongs to the system class $\mathcal{H}_{\textup{SRG}}(\mathcal{S})$. 

\begin{theorem}\label{lem:3}
Given a set $\mathcal{S} \subseteq \mathbb{C}$ as in \eqref{eq:set_complex_region}, and consider a feedback interconnection $[H_1,H_2]$. Suppose the following holds:
 \begin{enumerate}[label=(S\arabic*)]
        \item $H_1\in \mathcal{B}_\Delta$ and $\tau H_2 \in \mathcal{H}_{\textup{SRG}}(\mathcal{S})$ for all $\tau \in (0,1]$; \label{item_Thm3_S1}
        \item $[H_1, \tau H_2]$ is well-posed for all $\tau \in (0,1]$; \label{item_Thm3_S2}
        \item \label{item_Thm3_S3} there exists $r >0$ such that  
\begin{equation}\label{eq:cond0}
   \textup{dist}(\textup{SRG}^{-1}(-H_1), \mathcal{S})\geq  r.
   \end{equation} 
    \end{enumerate}
    Then, the interconnection $[H_1,H_2]$ is incrementally bounded.
\end{theorem}

\begin{proof}
     The proof follows directly from \cite[Corollary 1]{Cha25}. 
\end{proof}

When specializing Theorem~\ref{lem:3} to the full LTI setting, the notion of frequency-wise SRG \cite{Chen25cdc} is less conservative and often more useful than the operator-theoretic SRG in Definition~\ref{eq:SG}. The frequency-wise SRG of an LTI system $H$ (with $H(s)\in \mathcal{RH}_{\infty}^{n\times n}$ and its frequency response matrix $H(j\omega)$) is defined for all $\omega\in[-\infty,\infty]$ as  
\begin{equation}\label{eq:freq}
\begin{split}
   & \textup{SRG}_\omega (H):=\big\{\rho(v,H(j\omega)v)e^{\pm j\theta(v,H(j\omega)v)}\mid \\
&\qquad\qquad\qquad\qquad\qquad\qquad\qquad\qquad0\neq v\in \mathbb{C}^n \big\},
    \end{split}
\end{equation}
in which the norms and inner-product appearing in $\rho$ and $\theta$ are understood in the Euclidean sense. In the case where $H$ is single-input
single-output, the frequency-wise SRG recovers the Nyquist diagram \cite[Proposition 1]{Chen25cdc}.  The inverse frequency-wise SRG can be defined in an analogous manner to \eqref{eq:SGi}. Moreover,  given a set $\mathcal{S}$ and frequency $\omega \in [-\infty, \infty]$, the frequency-wise SRG system class is defined as
\begin{equation}\label{eq:FWSRG}
    \mathcal{H}^\omega_{\textup{SRG}}(\mathcal{S}):=\left\{H(s) \in \mathcal{RH}_{\infty}^{n\times n}\;\middle|\; \textup{SRG}_\omega(H)\subseteq \mathcal{S}\right\}.
\end{equation}
The following result provides frequency-wise SRG conditions for stability analysis of LTI feedback systems.

\begin{theorem}\label{th:fw}
 Given a set $\mathcal{S}\subseteq \mathbb{C}$ as in \eqref{eq:set_complex_region}, and consider a feedback interconnection $[H_1,H_2]$, where $H_1$ and $H_2$ are LTI systems. Suppose $H_1(s)\in \mathcal{RH}_{\infty}^{n\times n}$ and $\tau H_2(s) \in \mathcal{H}^\omega_{\textup{SRG}}(\mathcal{S})$ for all $\tau \in (0,1]$. If there exists $r>0$ such that
\begin{equation}
    \textup{dist}(\textup{SRG}_\omega^{-1}(-H_1)\cap \mathcal{S})\geq r \textup{ for all } \omega \in [-\infty, \infty],
\end{equation}
then the interconnection $[H_1,H_2]$ is incrementally bounded.
\end{theorem}
\begin{proof}
    The proof follows from \cite[Theorem 2]{Chen25cdc}.
\end{proof}
Later in Section~\ref{sec:converse}, we primarily work with frequency-wise SRGs to establish converse results for LTI systems. For the equivalence results in the following section, we use the more general operator-theoretic SRG framework to accommodate nonlinear feedback systems.

\section{IQC and SRG Equivalences}\label{sec:equivalence}
This section presents our first main result. We pose and address the following question: 
\begin{question}
Under which conditions are the disjunctive and conjunctive IQC conditions of Theorems~\ref{lem:02} and \ref{lem:2} equivalent to the SRG conditions in Theorem~\ref{lem:3}?  
\end{question}
Since the well-posedness conditions in item~\ref{item_Thm2_I2A} of Theorem~\ref{lem:02} and item~\ref{item_Thm2_I2B} of Theorem~\ref{lem:2} are identical to item~\ref{item_Thm3_S2} of Theorem~\ref{lem:3}, it remains to establish class equivalence and separation equivalence.

\subsection{Class Equivalence}
To establish conditions for class equivalences such that \ref{item_Thm2_I1A} $\iff$ \ref{item_Thm3_S1} and \ref{item_Thm2_I1B} $\iff$ \ref{item_Thm3_S1}, we first introduce a number of useful sets. 

Let $\mathbf{\Pi}$ be a collection of a finite number of distinct symmetric matrices $\Pi \in \mathbb{R}^{2\times 2}$ with $\det({\Pi})<0$. The collection $\mathbf{\Pi}$ serves as a key bridge between an IQC-based system class to an SRG-based one. Given $\mathbf{\Pi}$, denote the family of multipliers constructed from $\mathbf{\Pi}$ as
\begin{equation}\label{eq:MPI1}
    \mathbf{M}_{\mathbf{\Pi}} : = \left\{ M \in \mathbb{R}^{2n \times 2n} \;\middle|\; M = \Pi \otimes I, \: \Pi \in \mathbf{\Pi}\right\}.
\end{equation}
Clearly, $\mathbf{M}_{\mathbf{\Pi}}\subseteq \mathbf{M}$ with $\mathbf{M}$ as in \eqref{eq:set_multiplier_M}. The disjunctive and conjunctive complex sets in $\mathbb{C}$ generated from the collection $\mathbf{\Pi}$ are respectively defined as
\begin{equation}\label{eq:SPId}    {\mathcal{S}}^\cup_{\mathbf{\Pi}}:= \bigcup_{\Pi \in \mathbf{\Pi}}\left\{z \in \mathbb{C} \;\middle|\; \begin{bmatrix}
            1\\z
        \end{bmatrix}^*\Pi \begin{bmatrix}
            1\\z
        \end{bmatrix}\geq 0 \right\},
\end{equation}
and
\begin{equation}\label{eq:SPI}
    {\mathcal{S}}^\cap_{\mathbf{\Pi}}:= \bigcap_{\Pi \in \mathbf{\Pi}} \left\{ z\in \mathbb{C} \;\middle|\; \begin{bmatrix}
            1\\z
        \end{bmatrix}^*\Pi \begin{bmatrix}
            1\\z
        \end{bmatrix}\geq 0\right\}.
\end{equation}
Similarly, one can verify that ${\mathcal{S}}^\cup_{\mathbf{\Pi}}\subseteq \mathcal{S}$ and ${\mathcal{S}}^\cap_{\mathbf{\Pi}} \subseteq \mathcal{S}$, with $\mathcal{S}$ as in \eqref{eq:set_complex_region}. Using the above definitions, we obtain the next result.

\begin{lemma}\label{lem:eq} 
The following class equivalences hold:
\begin{enumerate}[label=\alph*)]
    \item $\mathcal{H}^\cup_{\textup{IQC}}(\mathbf{M}_{\mathbf{\Pi}}) = \mathcal{H}_{\textup{SRG}}(\mathcal{S}^\cup_{\mathbf{\Pi}})$;
    \item $\mathcal{H}^\cap_{\textup{IQC}}(\mathbf{M}_{\mathbf{\Pi}}) = \mathcal{H}_{\textup{SRG}}(\mathcal{S}^\cap_{\mathbf{\Pi}})$.
\end{enumerate}
\end{lemma}

\begin{proof}
$a)$: Follows directly from the construction of $\mathcal{S}^\cup_{\mathbf{\Pi}}$ and trajectory-wise application of \cite[Lemma 5]{Groot25}. 

   $ b)$: Follows from \cite[Lemma 5]{Groot25}, which establishes equivalence for $|\mathbf{\Pi}|=1$. The case $|\mathbf{\Pi}|>1$ can be obtained by applying \cite[Lemma 5]{Groot25} repeatedly. 
\end{proof}
The structure of the disjunctive and conjunctive sets $\mathcal{S}^\cup_{\mathbf{\Pi}}$ and $\mathcal{S}^\cap_{\mathbf{\Pi}}$ in \eqref{eq:SPId} and \eqref{eq:SPI} comes naturally for many relevant systems. For instance, gain-scheduled and switched systems may exhibit different dissipativity properties at different operating points or switching conditions \cite{Leith}, while mixed systems may satisfy different properties depending on the nature of the input \cite{EijCha,Eij26}. The SRG for these types of systems is then naturally contained in the disjunctive set $\mathcal{S}^\cup_{\mathbf{\Pi}}$. 

An example of a conjunctive characterization can be found in \cite{Groot25}, where the SRG of bounded LTI systems was shown to be characterized exactly by $\mathcal{S}^\cap_{\mathbf{\Pi}}$ for the family $$\mathbf{\Pi}= \left\{\sigma \begin{bmatrix}
    r^2-\lambda^2 & \lambda \\
    \lambda & -1
\end{bmatrix} \;\middle|\; \sigma =\left\{-1,1\right\}, \lambda \in \mathbb{R}, r>0\right\}.$$

Relevant conic regions, including tall ellipsoids considered in \cite{Baron} for bounding the SRG of LTI systems, can likewise be represented as intersections of infinitely many disks. This is formalized in the next lemma.

\begin{lemma}\label{lem:cone}
Define the conic region
\begin{equation}
    \mathcal{C}(Q) := \left\{ z \in \mathbb{C} \mid v(z)^\top Q v(z) \le 0 \right\},
\end{equation}
with the symmetric matrix $Q  \in \mathbb{R}^{3\times 3}$ of the form
$$
Q=
\begin{bmatrix}
q_{11} & 0 & q_{13} \\
0 & q_{22} & 0 \\
q_{13} & 0 & q_{33}
\end{bmatrix}, \textup{ where } q_{11}\geq q_{22} > 0, 
$$
and with $v(z) = \begin{bmatrix}\textup{Re}(z) & \textup{Im}(z) &1\end{bmatrix}^\top\in \mathbb{R}^3$. Then, 
$$
\mathcal{C}(Q)
=
\bigcap_{\lambda \in \mathbb{R}} \left\{ z \in \mathbb{C} \;\middle|\;
\begin{bmatrix}1 \\ z\end{bmatrix}^* \Pi_\lambda \begin{bmatrix}1 \\ z\end{bmatrix} \ge 0
\right\},
$$
in which
$$
\Pi_\lambda =
\begin{bmatrix}
\alpha \lambda^2 -q_{33} & -q_{13} -\alpha \lambda \\
-q_{13} - \alpha \lambda & -q_{22}
\end{bmatrix},
$$
and $\alpha:= q_{11} - q_{22}\geq 0$.
\end{lemma}

\begin{proof}
Writing $z = x + i y \in \mathbb{C}$ yields
\begin{equation*}
\begin{split}
v(z)^\top Q v(z)
&= q_{11} x^2 + q_{22} y^2 + 2q_{13} x + q_{33}\\
& =\alpha x^2+q_{22}|z|^2+2q_{13}x+q_{33}\leq 0.
\end{split}
\end{equation*}
Since $\alpha \geq 0$, we have $\alpha x^2 = \sup_{\lambda \in \mathbb{R}} \alpha (2\lambda x - \lambda^2)
$, from which it follows that
$$
\sup_{\lambda \in \mathbb{R}}
\left[
\alpha (2\lambda x - \lambda^2)
+ q_{22}|z|^2
+ 2q_{13} x
+ q_{33}
\right]
\le 0.
$$
This can equivalently be written as
$$
\alpha (2\lambda x - \lambda^2)
+ q_{22}|z|^2
+ 2q_{13} x
+ q_{33}
\le 0, \quad\forall \lambda \in \mathbb{R}.
$$
Using $x = \tfrac{1}{2}(z+z^*)$ and rearranging terms yields 
$$
\begin{bmatrix}1 \\ z\end{bmatrix}^*
\begin{bmatrix}
\alpha \lambda^2 -q_{33} & -q_{13} - \alpha \lambda \\
-q_{13} - \alpha \lambda & -q_{22}
\end{bmatrix}
\begin{bmatrix}1 \\ z\end{bmatrix}
\ge 0,\quad\forall \lambda \in \mathbb{R},
$$
which proves the result.
\end{proof}

We note that the conic regions in Lemma~\ref{lem:cone}, obtained as intersections of infinitely many disks, can be approximated arbitrarily closely by the intersection of a sufficiently large finite collection of disks; see also \cite{Groot25}. The result in Lemma~\ref{lem:cone} also highlights a practical distinction between the SRG and IQC frameworks: representing conic regions as infinitely many IQCs can lead to intractable problems in IQC analysis, and finite approximations may introduce additional conservatism. In contrast, SRGs are particularly well suited to handling such cases as they allow us to work with the conic region directly in the complex plane. 

\begin{figure}[htbp]
    \centering
\includegraphics[ width=.45\textwidth]{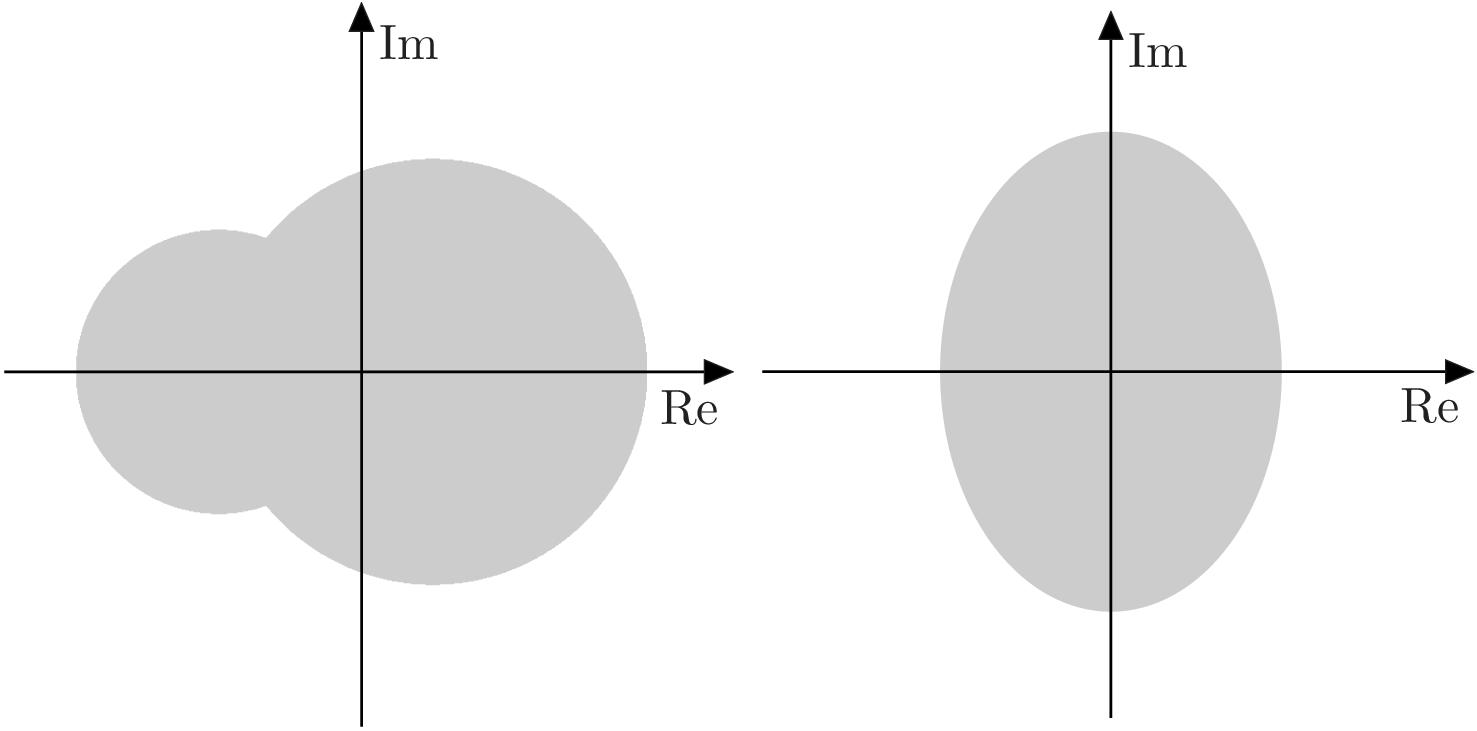}
    \caption{Left: disjunctive set (union of two disks). Right: conjunctive set (intersection of an infinite family of disks).}
    \label{fig:sets}
\end{figure}

Examples illustrating the disjunctive set $\mathcal{S}^\cup_{\mathbf{\Pi}}$ in \eqref{eq:SPId} and the conjunctive set $\mathcal{S}^\cap_{\mathbf{\Pi}}$ in \eqref{eq:SPI} are shown in Fig.~\ref{fig:sets}.

\subsection{Separation Equivalence}

We show the equivalence between the IQC and SRG separation conditions in this subsection. In particular, \ref{item_Thm2_I3A}~$\iff$~\ref{item_Thm3_S3} is established as is, and \ref{item_Thm2_I3B}~$\iff$~\ref{item_Thm3_S3} is established with an additional linearity assumption on the system. We start with the disjunctive case. 

{
\renewcommand{\thetheorem}{\number\numexpr\value{theorem}\relax.A}
\begin{theorem}\label{th:4A}
    Let $H\in \mathcal{B}_\Delta$ and a collection $\mathbf{\Pi}$ be given. Define $\mathbf{M}_{\mathbf{\Pi}}$ as in \eqref{eq:MPI1}. The following statements are equivalent:
    \begin{enumerate}[label=\arabic*.A),leftmargin=*]
        \item \label{item_Thm5A_1A} For all $M \in \mathbf{M}_{\mathbf{\Pi}}$, there exists $\epsilon >0$ such that
    \begin{equation}\label{eq:IQC_FDt}
      \int_{0}^\infty  \begin{bmatrix}
          \Delta y(t)\\ \Delta u(t)
        \end{bmatrix}^\top (M+\epsilon I) \begin{bmatrix}
            \Delta y(t)\\ \Delta u(t)
        \end{bmatrix}dt \leq 0, 
    \end{equation}
   holds for all $(y_1,u_1), (y_2,u_2)\in \mathcal{G}^{-1}(-H)$.
        \item \label{item_Thm5A_2A} There exists $r >0$ such that  
\begin{equation}\label{eq:cond00d}
   \textup{dist}(\textup{SRG}^{-1}(-H),  \mathcal{S}^\cup_{\mathbf{\Pi}})\geq  r,
   \end{equation}
   where $\mathcal{S}^\cup_{\mathbf{\Pi}}$ is as defined in \eqref{eq:SPId}.
    \end{enumerate}
\end{theorem}
}

\begin{proof}
\ref{item_Thm5A_1A} $\implies$ \ref{item_Thm5A_2A}. Partition each $\Pi \in \mathbf{\Pi}$ as $\Pi = \big[\begin{smallmatrix}
        a & b\\ b & c
    \end{smallmatrix}\big]$, $a,b,c\in \mathbb{R}$ such that each $M \in \mathbf{M}_{\mathbf{\Pi}}$ can be written as $M = \big[\begin{smallmatrix}
        a I & b I\\ b I & c I
    \end{smallmatrix}\big]$. Then, \eqref{eq:IQC_FDt} implies
    \begin{equation*}
    \begin{split}
      & \frac{1}{\|\Delta u\|^2}( a\|\Delta y\|^2+2b\langle \Delta y, \Delta u\rangle +c\|\Delta u\|^2) \\
       &  \qquad= a|z|^2 + 2b\textup{Re}(z)+c= \begin{bmatrix}
           z \\ 1
       \end{bmatrix}^*\Pi \begin{bmatrix}
           z \\ 1
       \end{bmatrix} \leq  -\epsilon \left\|\begin{bmatrix}
           z \\ 1
       \end{bmatrix}\right\|^2,
       \end{split}
    \end{equation*}
    where
    \begin{equation*}\begin{split}z
    & := \frac{\langle \Delta y, \Delta u\rangle}{\|\Delta u\|^2} \pm i \frac{\sqrt{\|\Delta u\|^2 \|\Delta y\|^2 - \langle \Delta y, \Delta u\rangle^2}}{\|\Delta u\|^2}.\end{split}\end{equation*}
By definition, $z \in \textup{SRG}(-H)$. It follows from $v := 1/\bar{z} = z/|z|^2 \in \textup{SRG}^{-1}(-H)$ and taking the intersection over all $\Pi \in \mathbf{\Pi}$ that 
\begin{equation*}
    \textup{SRG}^{-1}(-H) \subseteq \mathcal{S}^c_{\mathbf{\Pi}}:=\bigcap_{\Pi \in \mathbf{\Pi}}\left\{v \in \mathbb{C} \;\middle|\; \begin{bmatrix}
        1\\v
    \end{bmatrix}^*\Pi \begin{bmatrix}
        1\\v
    \end{bmatrix}<0\right\}.
\end{equation*}
Since $\mathcal{S}^\cup_{\mathbf{\Pi}} = \mathbb{C}\setminus \mathcal{S}^c_{\mathbf{\Pi}}$, the existence of $\epsilon >0$ implies that $\textup{SRG}^{-1}(-H)$ and $\mathcal{S}^\cup_{\mathbf{\Pi}}$ are uniformly strictly separated, so that \eqref{eq:cond00d} holds. 

\ref{item_Thm5A_2A} $\implies$ \ref{item_Thm5A_1A}. The implication follows by retracing the preceding steps in reverse order.    
\end{proof}

Next, we show the conjunctive case with an additional assumption on $H$ being LTI.

{
\renewcommand{\thetheorem}{\number\numexpr\value{theorem}\relax.B}
\addtocounter{theorem}{-1} % keep counter from advancing
\begin{theorem}\label{eq:sep_eq}
    Let $H\in \mathcal{L}$ and a collection $\mathbf{\Pi}$ be given; define $\mathbf{M}_{\mathbf{\Pi}}$ as in \eqref{eq:MPI1}. The following statements are equivalent:
    \begin{enumerate}[label=\arabic*.B),leftmargin=*]
 \item \label{item_Thm5B_1B} There exist scalars $\mu_M \geq 0$, not all zero, and $\epsilon >0$ such that 
 \begin{equation}\label{eq:IQC_0}
      \int_{0}^\infty  \begin{bmatrix}
          \Delta y(t)\\ \Delta u(t)
        \end{bmatrix}^\top (\sum_{M \in \mathbf{M}_{\mathbf{\Pi}}}\mu_M M+\epsilon I) \begin{bmatrix}
            \Delta y(t)\\ \Delta u(t)
        \end{bmatrix}dt \leq 0, 
    \end{equation}
     for all $(y_1,u_1), (y_2,u_2)\in \mathcal{G}^{-1}(-H)$. 
       \item \label{item_Thm5B_2B} There exists $r >0$ such that  
\begin{equation}\label{eq:cond00}
   \textup{dist}(\textup{SRG}^{-1}(-H), \mathcal{S}^\cap_{\mathbf{\Pi}})\geq  r,
   \end{equation}
   where $\mathcal{S}^\cap_{\mathbf{\Pi}}$ is as defined in \eqref{eq:SPI}.
    \end{enumerate}
\end{theorem}
}

\begin{remark}\normalfont
    Since $H$ is assumed to be LTI and bounded, the time-domain condition in \eqref{eq:IQC_0} can be reformulated via \cite[Proposition 4]{Jonsson01} as the equivalent frequency-domain condition:
    \begin{equation}\label{eq:IQC_FD1}
      \begin{bmatrix}
           -H(j\omega)\\ I
        \end{bmatrix}^* \sum_{M \in \mathbf{M}_{\mathbf{\Pi}}} \mu_M M \begin{bmatrix}
            -H(j\omega)\\I
        \end{bmatrix}<0,
    \end{equation}
    for all $\omega  \in [-\infty, \infty]$. This facilitates the numerical verification of \eqref{eq:IQC_0}. \hfill\scalebox{1.3}{$\diamond$}
\end{remark}

\begin{proof}[Proof of Theorem~\ref{eq:sep_eq}]
    \ref{item_Thm5B_1B}~$\implies$~\ref{item_Thm5B_2B}. We show the implication by contradiction.   Suppose that \ref{item_Thm5B_1B} holds, but \ref{item_Thm5B_2B} is false. Then $\textup{dist}(\textup{SRG}^{-1}(-H), \mathcal{S}_{\mathbf{\Pi}}^\cap)=0$. Note that this does not necessarily imply $\textup{SRG}^{-1}(-H)\cap \mathcal{S}^\cap_{\mathbf{\Pi}}\not = \emptyset$. Define the inflated set
    $$
    \mathcal{S}_{\mathbf{\Pi}_{\epsilon}}^\cap := \bigcap_{\Pi \in \mathbf{\Pi}} \left\{ z\in \mathbb{C} \;\middle|\; \begin{bmatrix}
            1\\z
        \end{bmatrix}^*\left(\Pi+\tilde{\epsilon}I\right) \begin{bmatrix}
            1\\z
        \end{bmatrix}\geq 0\right\},
    $$
   with $\tilde{\epsilon}>0$. Clearly, for all $\tilde{\epsilon}>0$ we have $\mathcal{S}_{\mathbf{\Pi}}^\cap \subseteq \mathcal{S}_{\mathbf{\Pi}_{\epsilon}}^\cap$. Since the sets $\textup{SRG}^{-1}(-H_1)$ and $\mathcal{S}_{\mathbf{\Pi}}^\cap$ are arbitrary close, it follows that $\textup{SRG}^{-1}(-H_1)\cap\mathcal{S}_{\mathbf{\Pi}_{\epsilon}}^\cap \neq \emptyset$ for any $\tilde{\epsilon} >0$. Choose $0<\tilde{\epsilon} < \epsilon /\sum_{M\in \mathbf{M}_{\mathbf{\Pi}}}\mu_M$, where $\epsilon$ and $\mu_M$ follow from $1.\textup{B})$. Pick $v \in \textup{SRG}^{-1}(-H)\cap \mathcal{S}^\cap_{\mathbf{\Pi}_\epsilon}$ and note that $|v|<\infty$. By construction of $\mathcal{S}^\cap_{\mathbf{\Pi}_\epsilon}$, we have that
    \begin{equation}\label{eq:iq1}
        \begin{bmatrix}
            1 \\ v
        \end{bmatrix}^* \left(\sum_{M \in \mathbf{M}_{\mathbf{\Pi}}} \mu_M (M+\tilde{\epsilon} I)\right) \begin{bmatrix}
            1 \\ v
        \end{bmatrix} \geq 0,
    \end{equation}
    for any $\mu_M \geq 0$. Let $\mu_M$ follow from \ref{item_Thm5B_1B}. Use the map 
    \begin{equation*}\begin{split}v
    &= \frac{\langle \Delta y, \Delta u\rangle}{\|\Delta y\|^2} \pm i \frac{\sqrt{\|\Delta u\|^2 \|\Delta y\|^2 - \langle \Delta y, \Delta u\rangle^2}}{\|\Delta y\|^2}\end{split}\end{equation*}
    with $(y_1, u_1), (y_2,u_2) \in \mathcal{G}^{-1}(-H)$, $\|\Delta y\|\neq 0$, and use the partition $\Pi = \big[\begin{smallmatrix}
        a & b\\ b & c
    \end{smallmatrix}\big]$, $a,b,c\in \mathbb{R}$ to find that \eqref{eq:iq1} implies
    \begin{equation*}
    \begin{split}
       & 0\leq \sum_{M \in \mathbf{M}_{\mathbf{\Pi}}}\mu_M (a + 2 b \textup{Re}(v) + c|v|^2+\tilde{\epsilon}(1+|v|^2)) \\
        &=\sum_{M \in \mathbf{M}_{\mathbf{\Pi}}}\frac{\mu_M}{\|\Delta y\|^2} (a \|\Delta y\|^2 + 2 b \langle \Delta y, \Delta u\rangle + c \|\Delta u\|^2) \\
        & \qquad \qquad + \sum_{M \in \mathbf{M}_{\mathbf{\Pi}}}\frac{\mu_M}{\|\Delta y\|^2}(\tilde{\epsilon}(\|\Delta y\|^2+\|\Delta u\|^2)\\
        & = \frac{1}{\|\Delta y\|^2} \int_{0}^\infty \begin{bmatrix}
            \Delta y \\\Delta u
        \end{bmatrix} ^\top \left(\sum_{M \in \mathbf{M}_{\mathbf{\Pi}}} \mu_M (M+\tilde{\epsilon}I)\right)\begin{bmatrix}
            \Delta y \\\Delta u
        \end{bmatrix} dt \\
        &< \frac{1}{\|\Delta y\|^2} \int_{0}^\infty \begin{bmatrix}
            \Delta y \\\Delta u
        \end{bmatrix} ^\top \left((\sum_{M \in \mathbf{M}_{\mathbf{\Pi}}} \mu_M M+{\epsilon}I)\right)\begin{bmatrix}
            \Delta y \\\Delta u
        \end{bmatrix} dt,
        \end{split}
    \end{equation*}
    where in the last inequality we used $\tilde{\epsilon}\sum_{M\in \mathbf{M}_{\mathbf{\Pi}}}\mu_M < {\epsilon}$. This, however, directly contradicts \eqref{eq:IQC_0}.   
   
\ref{item_Thm5B_2B} $\implies$ \ref{item_Thm5B_1B}. 
Suppose \ref{item_Thm5B_2B} holds. By uniform separation of
$\mathrm{SRG}^{-1}(-H)$ and $\mathcal{S}^\cap_{\mathbf{\Pi}}$, there exists $\epsilon >0$ such that
\begin{equation}
\begin{split}
&\textup{SRG}^{-1}(-H)
\subseteq (\mathbb{C} \setminus \mathcal{S}^\cap_{\mathbf{\Pi}_\epsilon}) \\
& \qquad \quad\quad= \bigcup_{\Pi\in\mathbf{\Pi}}
\left\{
v\in\mathbb{C}\,\middle|\,
\begin{bmatrix}1\\ v\end{bmatrix}^*
\Pi_\epsilon
\begin{bmatrix}1\\ v\end{bmatrix}
\leq 0
\right\},
\end{split}
\end{equation}
where $\Pi_\epsilon := \Pi+\epsilon I$. Using the definition of the inverse SRG and the coordinate transformation $z = v^{-1}$, this implies
\begin{equation}\label{eq:SRG_v1}
\begin{split}
& \textup{SRG}(-H)
\subseteq \bigcup_{\Pi\in\mathbf{\Pi}}
\left(\left\{
v\in\mathbb{C}\,\middle|\,
\begin{bmatrix}1\\ v\end{bmatrix}^*
\Pi_\epsilon
\begin{bmatrix}1\\ v\end{bmatrix}
\leq 0
\right\}\right)^{-1} \\
& \qquad \qquad \:\:=\bigcup_{\Pi\in\mathbf{\Pi}}
\left\{
z\in\mathbb{C}\,\middle|\,
\begin{bmatrix}z\\ 1\end{bmatrix}^*
\Pi_\epsilon\begin{bmatrix}z\\ 1\end{bmatrix}
\leq 0
\right\}.
\end{split}
\end{equation}
Consider the Beltrami-Klein mapping $g: \mathbb{C} \to \mathbb{D}$ in \cite[Section~V.A]{Chaffey23} and let
\begin{subequations}
\begin{align}\label{eq:BK_mapping_LTI}
    W_{\textup{BK}}(H) &:= g(\textup{SRG}(-H)),\\
    S_{\textup{BK}}(\Pi) &: = g\left(\left\{ z\in\mathbb{C}\,\middle|\,
\begin{bmatrix}z\\ 1\end{bmatrix}^*
\Pi_\epsilon\begin{bmatrix}z\\ 1\end{bmatrix}
\leq 0 \right\}\right).
\end{align}
\end{subequations}
Due to \eqref{eq:SRG_v1} it follows that
\begin{equation}\label{eq:Upi}
    W_{\textup{BK}}(H) \subseteq U_{\mathbf{\Pi}}:=\bigcup_{\Pi \in \mathbf{\Pi}} S_{\textup{BK}}(\Pi).
\end{equation}
Since $H$ is LTI, it follows from \cite{Chaffey23} that $W_{\textup{BK}}(H)$ is a convex set. Moreover, from the assumption that $\det(\Pi)<0$ it follows that each region on the right-hand side of \eqref{eq:SRG_v1} that is defined by $\Pi$ represents either the interior of a disk, exterior of a disk, or a half plane. As a consequence of the Beltrami-Klein mapping that maps circles and lines in $\mathbb{C}$ to lines in $\mathbb{D}$, each $S_{\textup{BK}}(\Pi)$ is a convex region in $\mathbb{D}$ generated by the intersection of a half-plane $\mathcal{H}$ and the unit disk, i.e., $S_{\textup{BK}}(\Pi) = \mathbb{D} \cap \mathcal{H}(\Pi)$. Hence, the union of these regions $U_\mathbf{\Pi}$ in \eqref{eq:Upi} is non-convex. Consequently, by De Morgans law, $\mathbb{D} \setminus U_{\mathbf{\Pi}} = \cap_{\Pi \in \mathbf{\Pi}} (\mathbb{D}\setminus  S_{\textup{BK}}(\Pi))$ is convex. By virtue of the separating hyperplane theorem \cite{Boyd}, there exists a line $\ell$ in $U_{\mathbf{\Pi}}$ separating $W_{\textup{BK}(H)}$ and $\overline{\mathbb{D} \setminus U_{\mathbf{\Pi}}}$. This line induces a new convex region in $\mathbb{D}$, denoted $V$, so that $W_{\textup{BK}}(H) \subseteq V \subseteq U_{\mathbf{\Pi}}$. An exemplary situation is illustrated in Fig.~\ref{fig:WBK}.

\begin{figure}[htbp]
    \centering
\includegraphics[ width=.35\textwidth]{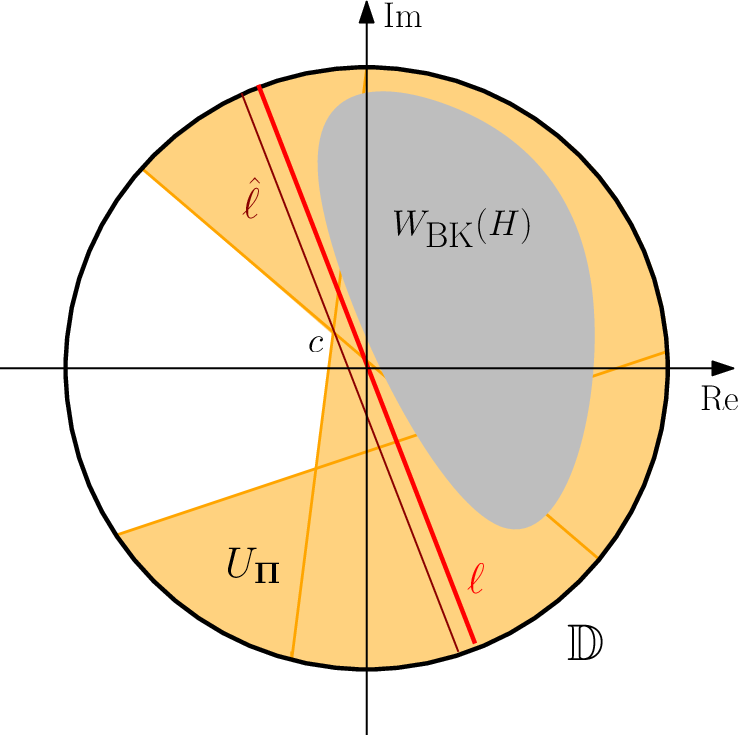}
    \caption{Illustration of the relevant regions in $\mathbb{D}$: (a) the region $W_{\textup{BK}}(H)$ in grey, (b) the region $U_{\mathbf{\Pi}}$ formed as the union of three sets $S_{\textup{BK}}(\Pi)$ in orange, (c) the separating line $\ell$ in red, and $(c)$ the shifted separating line $\hat{\ell}$ in dark red. The regions $V$ and $\hat{V}$ are the half-planes to the right of $\ell$ and $\hat{\ell}$, respectively.}
    \label{fig:WBK}
\end{figure}

We next show that there exist $\Pi_i, \Pi_j \in \mathbf{\Pi}$ such that
\begin{equation}\label{eq:VSS}
    V \subseteq S_{\textup{BK}}(\Pi_i) \cup S_{\textup{BK}}(\Pi_j).
\end{equation}
Since $\ell \subseteq V$, we find $\ell \subseteq U_{\mathbf{\Pi}}$. For each $\Pi \in \mathbf{\Pi}$, the intersection $\ell \cap S_{\textup{BK}}(\Pi)$ has three possibilities: i) the entire line $\ell$, ii) a half line in $\ell$, or iii) empty.  

Case i): If $\ell \cap S_{\textup{BK}}(\Pi) = \ell$ for some $\Pi \in \mathbf{\Pi}$, then either $V \subseteq S_{\textup{BK}}(\Pi)$ or $V\not\subset S_{\textup{BK}}(\Pi)$. In the first case, we are done. In the second case, since $\ell \subseteq V$ we have $V\cup S_{\textup{BK}}(\Pi) = \mathbb{D}$, and, therefore, $U_{\mathbf{\Pi}} = \mathbb{D}$ (i.e., $V$ and $S_{\textup{BK}}(\Pi)$ are complementary regions in $\mathbb{D}$). Since $V \subseteq U_{\mathbf{\Pi}}$, in that case we can simply pick a $\Pi \in \mathbf{\Pi}$ so that $V = S_{\textup{BK}}(\Pi)$. 

Case ii): Identifying $\ell$ with $\mathbb{R}$, each nonempty $ S_{\textup{BK}}(\Pi) \cap \ell$ is of the form
$(-\infty,a_j]$ or $[a_i,\infty)$. Let $(-\infty,a_j]$ be the left‑half‑line with maximal endpoint $a_j$, and let $[a_i,\infty)$ be the right‑half‑line with minimal endpoint $a_i$. Then
\begin{equation*}
\ell = (-\infty,a_j] \cup [a_i,\infty)
= (S_{\textup{BK}}(\Pi_j) \cap \ell) \cup (S_{\textup{BK}}(\Pi_i) \cap \ell).
\end{equation*}
Then, either $V \subseteq S_{\textup{BK}}(\Pi_i) \cup S_{\textup{BK}}(\Pi_j)$ must be true, or $V\cup S_{\textup{BK}}(\Pi_i) \cup S_{\textup{BK}}(\Pi_j) = \mathbb{D}$, and we end up with the same situation as in Case i). 

Case iii): The case that the intersection of $\ell$ and $S_{\textup{BK}}(\Pi)$ for \emph{all} $\Pi \in \mathbf{\Pi}$ is empty can be ruled out, since $\ell \subseteq V \subseteq U_{\mathbf{\Pi}}$, and thus there must exist at least one $\Pi \in \mathbf{\Pi}$ for which $\ell$ and $S_{\textup{BK}}(\Pi)$ intersect. 

In conclusion: \eqref{eq:VSS} holds for some $\Pi_i, \Pi_j \in \mathbf{\Pi}$. This is also shown in Fig.~\ref{fig:WBK}.

Denote $b_i = \partial S_{\textup{BK}}(\Pi_i)$ and $b_j= \partial S_{\textup{BK}}(\Pi_j)$ as the straight boundary lines from $S_{\textup{BK}}(\Pi_i)$ and $S_{\textup{BK}}(\Pi_j)$, which are chords of the unit disk (see the dark orange lines in Fig.~\ref{fig:WBK}). Their intersection $c:=b_i \cap b_j$ is a single point in the interior of $\mathbb{D}$. Define the translated line $\hat{\ell}$ that is parallel to $\ell$ and passes through the point $c$ (see Fig.~\ref{fig:WBK}). Note that $\hat{\ell}$ defines a region $\hat{V}$ such that $V\subseteq \hat{V} \subseteq  S_{\textup{BK}}(\Pi_i) \cup S_{\textup{BK}}(\Pi_j)$. Using the inverse Beltrami-Klein mapping $f:\mathbb{D} \rightrightarrows \mathbb{C}$ in \cite[Section~V.A]{Chaffey23} to $b_i, b_j$, and $\hat{\ell}$ maps these straight lines into the circles or lines
\begin{equation*}
    \begin{bmatrix}
        v \\ 1
    \end{bmatrix}^* (\Pi_k+\epsilon I)  \begin{bmatrix}
        v \\ 1
    \end{bmatrix} = 0, \textup{ and }  \begin{bmatrix}
        v \\ 1
    \end{bmatrix}^* \Psi  \begin{bmatrix}
        v \\ 1
    \end{bmatrix} = 0,
\end{equation*}
with $k = i,j$, where $\Pi_k$ follow from $b_i,b_j$ and $\Psi$ follows from $\hat{\ell}$. Note that all circles share the same intersection points $[p, p^*]$ determined by $c$, and the circle associated with $\Psi$ is contained in the union of the disks generated by $\Pi_{k}+\epsilon I$. By the theory of coaxial pencil of circles, a circle passing through the points $[p, p^*]$ and that is contained in the union of the two generating circles has the implicit form
\begin{equation}
    \begin{bmatrix}
        v\\1
    \end{bmatrix}^* \Psi \begin{bmatrix}
        v \\1
    \end{bmatrix} = 0 \textup{ with } \Psi= \sum_{k=i,j} \alpha_k(\Pi_k+\epsilon I),
\end{equation}
with $\alpha_k \geq 0$. Since the disk $D_{\Psi}$ generated by $\Psi$ is a subset of those generated by $\Pi_k+\epsilon I$, and $\textup{SRG}(-H) \subseteq f(W_{\textup{BK}}(H)\subseteq D_{\Psi}$, it must hold that
\begin{equation}
    \begin{bmatrix}
        v \\ 1
    \end{bmatrix}^*  \sum_{k=i,j} \alpha_k(\Pi_k+\epsilon I)\begin{bmatrix}
        v \\ 1
    \end{bmatrix} \leq 0, \:\:\forall v \in \textup{SRG}(-H). 
\end{equation}
Setting
\begin{equation*}\begin{split}v
    &= \frac{\langle \Delta u, \Delta y\rangle}{\|\Delta u\|^2} \pm i \frac{\sqrt{\|\Delta u\|^2 \|\Delta y\|^2 - \langle \Delta y, \Delta u\rangle^2}}{\|\Delta u\|^2}\end{split}\end{equation*}
    with $(u_1, y_1), (u_2,y_2) \in \mathcal{G}(-H)$ establishes that \eqref{eq:IQC_0} holds with $\mu_i = \alpha_i\geq 0$, $\mu_j = \alpha_j \geq 0$ and $\mu_k = 0$ for $k \neq i,j$.
\end{proof}

The only step in the proof of Theorem~\ref{eq:sep_eq} that relies critically on the assumption that $H$ is LTI is the establishment of convexity of $W_{\textup{BK}}(H)$ in \eqref{eq:BK_mapping_LTI}, which then enables the use of a hyperplane separation argument (recall also Fig.~\ref{fig:WBK}). In general, convexity of $W_{\textup{BK}}(H)$ is not known to hold for arbitrary nonlinear systems $H$. However, it does hold for certain \emph{classes} of nonlinearities, such as static sector-bounded nonlinearities, suggesting that the LTI assumption in Theorem~\ref{eq:sep_eq} could potentially be relaxed and replaced by an assumption on the system class. We leave such an extension for future work.  

When the set $\mathcal{S}^\cap_\mathbf{\Pi}$ is characterized by a single quadratic inequality, i.e., the cardinality $|\mathbf{\Pi}|=1$, the LTI assumption can also be removed. Indeed, in this case we directly find $W_{\textup{BK}}(H) \subseteq S_{\textup{BK}}(\Pi)$ so that the remaining steps in the proof of Theorem~\ref{eq:sep_eq} after \eqref{eq:VSS} follow immediately. Alternatively, this conclusion follows from Theorem~\ref{th:4A}. Indeed, for $|\mathbf{\Pi}|=1$ we have $\mathcal{S}^\cap_{\mathbf{\Pi}} = \mathcal{S}^\cup_{\mathbf{\Pi}}$, and Theorem~\ref{th:4A} applies without requiring $H$ to be LTI. 

\subsection{Equivalence of Feedback Stability Results}
\begin{figure}[ht]
\centering
$$
\begin{tikzcd}[column sep=.1em, row sep=4em, arrows=Leftrightarrow]
& \begin{array}{c}
\textup{Theorem~\ref{lem:3}}\\
\textup{(SRG)}
\end{array} \arrow[dl, "\large \substack{\mathbf{M}=\mathbf{M}_{\mathbf{\Pi}} \\[2.5pt]\mathcal{S}=\mathcal{S}^\cup_{\mathbf{\Pi}}}" swap, , pos=0.5] \arrow[dr, "\large \substack{\mathbf{M}=\mathbf{M}_{\mathbf{\Pi}} \\[2.5pt] \mathcal{S}=\mathcal{S}^\cap_{\mathbf{\Pi}}\\[3.5pt]H_1 \:\:\textup{LTI}}", pos=0.7] & \\
\begin{array}{c}
\textup{Theorem~\ref{lem:02}}\\
\textup{(Disj. IQC)}
\end{array}
\arrow[rr, Rightarrow, yshift=4pt]
    \arrow[rr, "\large \substack{\\[3pt]\mathbf{M}=\mathbf{M}_{\mathbf{\Pi}}\\[2pt]|\Pi|=1}" below,yshift=-4pt] & & \begin{array}{c}
\textup{Theorem~\ref{lem:2}}\\
\textup{(Conj. IQC)}
\end{array}
\end{tikzcd}
$$
\caption{Equivalence relationships between IQC- and SRG-based feedback theorems under different sets of assumptions.}
\label{fig:equivalence}
\end{figure}
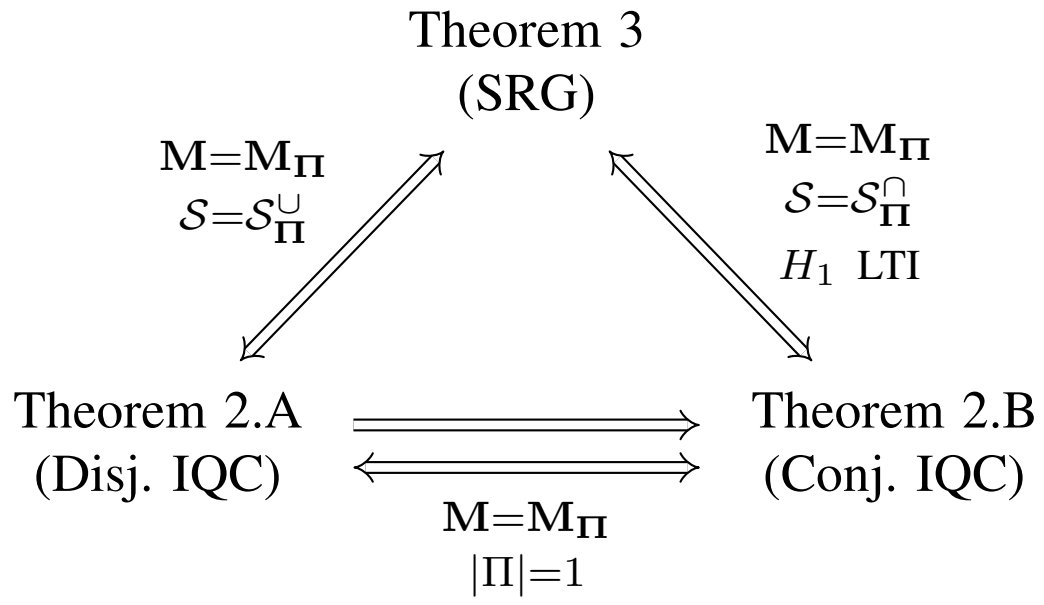

By combining the above class equivalence in Lemma~\ref{lem:eq} and separation equivalence in Theorems~\ref{th:4A} and \ref{eq:sep_eq}, we obtain the following main conclusions and answer to the posed question:
\begin{enumerate}
    \item When restricting the sets $\mathbf{M}=\mathbf{M}_{\mathbf{\Pi}}$ and $\mathcal{S}=\mathcal{S}^\cup_{\mathbf{\Pi}}$, it directly follows from Lemma~\ref{lem:eq} and Theorem~\ref{th:4A} that
    $$
    \textup{Theorem~\ref{lem:02}}\:\Longleftrightarrow\:\textup{Theorem~\ref{lem:3}}.
    $$

    \item When restricting the sets $\mathbf{M}=\mathbf{M}_{\mathbf{\Pi}}$ and $\mathcal{S}=\mathcal{S}^\cap_{\mathbf{\Pi}}$, and with the additional assumption that $H_1$ is LTI, it follows directly from Lemma~\ref{lem:eq} and Theorem~\ref{eq:sep_eq} that
     $$
\textup{Theorem~\ref{lem:2}}\:\Longleftrightarrow\:\textup{Theorem~\ref{lem:3}}.
    $$

    \item Since $\mathcal{H}^\cap_{\textup{IQC}}(\mathbf{M}) \subseteq \mathcal{H}^\cup_{\textup{IQC}}(\mathbf{M})$ and \ref{item_Thm2_I3A}~$\implies$~\ref{item_Thm2_I3B}, it follows that, in general, $$
    \textup{{Theorem~\ref{lem:02}}}\:\Rightarrow \: \textup{Theorem~\ref{lem:2}}.$$
   \item When restricting $\mathbf{M}=\mathbf{M}_{\mathbf{\Pi}}$ with $|\mathbf{\Pi}|=1$, and $\mathcal{S}=\mathcal{S}^\cup_{\mathbf{\Pi}}=\mathcal{S}^\cap_{\mathbf{\Pi}}$ where the latter equality results from $|\mathbf{\Pi}|=1$, it follows from Theorem~\ref{th:4A} and item 1) above that $$
    \textup{Theorem~\ref{lem:02} }\:\Longleftrightarrow \:\textup{Theorem~\ref{lem:2}}\:\Longleftrightarrow\:\textup{Theorem~\ref{lem:3}}.
    $$
\end{enumerate}
The equivalence relationships are summarized in Fig.~\ref{fig:equivalence}.

\begin{remark} \normalfont The established equivalences allow advantages of one framework to compensate for limitations of the other. For example, the conjunctive IQC theorem (Theorem~\ref{lem:2}) requires the search for a set of parameters $\mu_M$, whose number grows with the cardinality of $\mathbf{\Pi}$. For conic regions discussed in Lemma~\ref{lem:cone}, which are characterized by a high (infinite) number of matrices $\Pi$, this search quickly becomes intractable. In contrast, the equivalent SRG theorem (Theorem~\ref{lem:3}) enables a direct graphical verification, thereby avoiding the need to search over $\mu_M$. \hfill\scalebox{1.5}{$\diamond$}
\end{remark}

We conclude this section with some final remarks and observations.
\begin{enumerate}
    \item The estblished equivalences rely fundamentally on the quadratic structure of the sets $\mathbf{M}_{\mathbf{\Pi}}$, $\mathcal{S}^\cup_{\mathbf{\Pi}}$, and $\mathcal{S}^\cap_{\mathbf{\Pi}}$, all specified by a collection of $\mathbf{\Pi}$ in $\mathbb{R}^{2\times 2}$. For more general choices of $\mathbf{M}$ and $\mathcal{S}$, equivalence may not hold, and the IQC and SRG frameworks yield different results.
\item The multiplier class $\mathbf{M}$ under consideration is \emph{static}. The general IQC framework admits \emph{dynamic} multipliers using $\mathcal{L}_\infty$ functions \cite{Megretski
}. It is worth noting that our results can be directly adapted to the setting where sets of dynamic multipliers admit spectral factorizations with shared dynamics, see \cite{Seiler15}. 
\item The above results do not fundamentally rely on the ``soft'' nature of the SRGs and IQCs, i.e., on their action on the $\mathcal{L}_2$-space. Consequently, the results may be extended directly to the ``hard'' setting, where IQCs and SRGs act on the extended $\mathcal{L}_{2}$-space \cite{Khong21IQC, Su2, Chen26}. 
\end{enumerate}

\section{Converse Theorems}\label{sec:converse}
 
Equivalence of IQC- and SRG-based conditions enables results established in one framework to be translated directly to the other. This is particularly useful for deriving converse results, which play an important role in robustness analysis. This section presents our second set of main results in the form of general converse SRG and IQC theorems.

\subsection{General Converse Result}
We start with a general converse result for nonlinear systems based on the \emph{operator SRG}. To this end, we introduce the notion of SRG-realizability. Let $\mathcal{M}\subseteq\mathcal{B}_\Delta$ denote a structural class of causal and incrementally bounded systems.

\begin{definition}[SRG-realizability]\label{def:rel_gen}
  Let $\mathcal{S}\subseteq\mathbb{C}$ and $\mathcal{M}\subseteq\mathcal{B}_\Delta$.
We say that $\mathcal{S}$ is \emph{$\mathrm{SRG}(\mathcal{M})$-realizable} if, for every $M\in\mathcal{M}$ and some $u_1,u_2\in\mathcal{L}_2$, $u_1-u_2 \neq 0$ such that $(Mu_1-Mu_2,u_1-u_2)\mapsto z \in \mathcal{S}$, there exists a system $H \in \mathcal{B}_\Delta$ satisfying $u_1-u_2 = H(M)u_1-H(M)u_2$ and $\textup{SRG}(H) \subseteq \mathcal{S}$. When $\mathcal{M}=\mathcal{B}_\Delta$ we simply say that $\mathcal{S}$ is \emph{$\textup{SRG}$-realizable}. 
\end{definition}

SRG-realizability means that every SRG point contained in $\mathcal{S}$ admits a causal and bounded system realization whose entire SRG remains in $\mathcal{S}$. This property enables the construction of destabilizing systems in converse results by extending a critical input to a full system realization. In fact, it is this key mechanism that is used in classical LTI converse results such as the small-gain theorem \cite[Chapter~9]{ZDG}.

Before formulating our general converse result, we need some additional set properties. For $\epsilon >0$, we define an $\epsilon$-deflation of a set $\mathcal{S} \subseteq \mathbb{C}$ as
\begin{equation}
    \mathcal{S}^\epsilon := \left\{z \in \mathbb{C}\mid \textup{dist}(z,\mathbb{C}\setminus \mathcal{S})\geq \epsilon\right\}.
\end{equation}
For $\delta >0$, we define the $\delta$-inflation of a set
$\mathcal{S} \subseteq \mathbb{C}$ as
\begin{equation}
    \mathcal{S}_\delta := \left\{z \in \mathbb{C}\mid \textup{dist}(z,\mathcal{S})<\delta\right\}.
\end{equation}
Note that $\mathcal{S}^\epsilon \subseteq \mathcal{S} \subseteq \mathcal{S}_\delta$. The well-known notion of a star-shaped set is stated below.

\begin{definition}[Star-shaped set]\label{def:star}
   A set $\mathcal{S}\subseteq\mathbb{C}$ is said to be \emph{star-shaped with respect to the ball},
  $B(c,r)\subseteq \mathcal{S}$, if $(1-t)\,b+t\,s\in\mathcal{S}$ for all $b\in B(c,r)$, $s\in\mathcal{S}$, and $t\in[0,1]$.
  \end{definition}

  Many relevant disjunctive and conjunctive sets are star-shaped with respect to a ball, as we show later in this section. Now we are ready to present the following converse result. 

\begin{theorem}\label{th:general}
    Let $\mathcal{S}_1 \subseteq \mathbb{C}$ with $0\in \mathcal{S}_1$, and $\mathcal{S}_2 = (\mathbb{C}\setminus \mathcal{S}_1)^{-1}$. Let $\mathcal{M}\subseteq \mathcal{B}_\Delta$ and $\delta_0>0$. Suppose the following holds:
    \begin{enumerate}
        \item $-H_1\in \mathcal{M}$;
        \item $\tau H_2 \in \mathcal{H}_{\textup{SRG}}(\mathcal{S}_2)$ for all $\tau \in (0,1]$;
        \item$\mathcal{S}_2$ is star-shaped with respect $B(c,r)\subseteq\mathcal{S}_2$;
        \item $\mathcal{S}_{2,\delta}$ is $\textup{SRG}(\mathcal{M})$-realizable for all $\delta \in (0,\delta_0]$;
        \item $[H_1,\tau H_2]$ is well-posed for all $\tau \in (0,1]$.
    \end{enumerate}
    Then
    \begin{align}\label{eq: feedback_uniform_gain}
         \sup_{H_2 \in \mathcal{H}_{\textup{SRG}}(\mathcal{S}_2)}\|F_{[H_1,H_2]}\|_\Delta <\infty  
    \end{align}
    if and only if {there exists $\epsilon >0$ such that $-H_1 \in \mathcal{H}_{\textup{SRG}}(\mathcal{S}_1^\epsilon)$}.
\end{theorem}

\begin{proof}
    \emph{Sufficiency} follows directly from the SRG-based feedback stability theorem (Theorem~\ref{lem:3}). Indeed, $H_2 \in \mathcal{H}_\textup{SRG}(\mathcal{S}_2)$ implies $\textup{SRG}(H_2) \subseteq \mathcal{S}_2$. Since $-H_1 \in \mathcal{H}_{\textup{SRG}}(\mathcal{S}^\epsilon_1) \subseteq \mathcal{H}_{\textup{SRG}}((\mathbb{C}\setminus \mathcal{S}_2)^{-1})$ it follows that $\textup{SRG}^{-1}(-H_1)$ and $\mathcal{S}_2$ are strictly separated. Hence, there exists $r>0$, depending on $\epsilon$, so that $\textup{dist}(\textup{SRG}(-H_1)^{-1}, \mathcal{S}_2)\geq r$. As the well-posedness and separation condition hold for all $\tau H_2 \in \mathcal{H}_{\textup{SRG}}(\mathcal{S}_2)$, $\tau \in (0,1]$, Theorem~\ref{lem:3} yields $\|(I+H_2H_1)^{-1}\|_\Delta<\tfrac{1}{r}<\infty$ for all $H_2 \in \mathcal{H}_{\textup{SRG}}(\mathcal{S}_2)$.
    
To show \emph{necessity}, we proceed by contradiction. Suppose that there does not exist an $\epsilon>0$ so that $-H_1 \in \mathcal{H}_{\textup{SRG}}(\mathcal{S}^\epsilon_1)$. Then, $\textup{dist}(\textup{SRG}^{-1}(-H_1), \mathcal{S}_2)=0$. As such, it follows that $\textup{SRG}^{-1}(-H_1)\cap \mathcal{S}_{2,\delta} \neq \emptyset$ for all $\delta >0$. 

Suppose the pair $(u_1-u_2,-H_1u_1+H_1u_2)$ with $u_1,u_2 \in \mathcal{L}_2$ and $u_1-u_2\neq 0$ maps to a point $z_0 \in \mathcal{S}_{2,\delta}$. By $\mathrm{SRG}(\mathcal{M})$-realizability of $\mathcal{S}_{2,\delta}$ in Definition~\ref{def:rel_gen} and since $\mathcal{S}_{2,\delta}$ is bounded (as $0 \in \mathcal{S}_1$), there exists an incrementally bounded system $\tilde{H}_2\in\mathcal{H}_{\mathrm{SRG}}(\mathcal{S}_{2,\delta})$ such that $\tilde{H}_2(M)u_1-\tilde{H}_2(M)u_2 -(u_1-u_2) =0$. Now construct ${H}_2 = \tau_\delta \tilde{H}_2+(1-\tau_\delta)c I$ with $\tau_\delta := r/(r+\delta)$. We then divide the proof into two steps. 

The first step is to derive a contradiction with the uniform incremental gain hypothesis in \eqref{eq: feedback_uniform_gain}. To see this, we have 
\begin{equation*}
\begin{split}
    &\|{H}_2(M)u_1-{H}_2(M)u_2 - (u_1-u_2)\| \\
    &\qquad\qquad=\|(\tau_\delta-1)(u_1-u_2-c(Mu_1-Mu_2))\| \\
      &\qquad\qquad \leq |\tau_\delta -1|(1+|c|\gamma_M)\|u_1-u_2\|,
    \end{split}
\end{equation*}
where $\gamma_M =\|M\|_\Delta<\infty$. Substituting $M=-H_1$ gives
$$
\frac{\|u_1-u_2 -({H}_2(-H_1)u_1-{H}_2(-H_1)u_2)\|}{\|u_1-u_2\|}\leq|\tau_\delta -1|k,
$$
for $u_1-u_2 \neq 0$, where $k = 1+|c|\gamma_M$. By the feedback equation \eqref{eq:system}, it follows that $w_1-w_2 = u_1-u_2-({H}_2(-H_1)u_1-{H}_2(-H_1)u_2)$. Therefore, the ratio
$$
\frac{\|u_1-u_2\|}{\|w_1-w_2\|}\geq \frac{1}{k|\tau_\delta -1|}
$$
can be made arbitrary large by taking $\delta \to 0$, which contradicts the gain hypothesis.

The second step is to show that ${H}_2 \in \mathcal{H}_{\textup{SRG}}(\mathcal{S}_2)$, as required in \eqref{eq: feedback_uniform_gain}. First, note that $c+\tau_\delta(\mathcal{S}_{2,\delta}-c)\subseteq\mathcal{S}_2$. Indeed, let
  $z\in\mathcal{S}_{2,\delta}$ and pick $s\in\mathcal{S}_2$ with $|z-s|<\delta$. Then
  $b:=c+\tfrac{r}{\delta}(z-s)\in B(c,r)$ and
  $c+\tau_\delta(z-c)=\tau_\delta s+(1-\tau_\delta)b\in\mathcal{S}_2$, since by assumption $\mathcal{S}_2$ is star-shaped
  with respect to $B(c,r)$. Second, ${H}_2=\tau_\delta \tilde{H}_2+(1-\tau_\delta)c I$ is causal with
  finite incremental gain. It follows from the SRG composition rules (see, e.g., \cite[Proposition 5]{Chaffey23}) that $\textup{SRG}({H}_2) = c(1-\tau_\delta)+\tau_\delta \textup{SRG}(H_2) \subseteq c+\tau_\delta (\mathcal{S}_{2,\delta}-c)\subseteq \mathcal{S}_2$. Hence, $H_2 \in \mathcal{H}_{\textup{SRG}}(\mathcal{S}_2)$.
  
  All in all, there exists no finite incremental feedback gain over all $H_2 \in \mathcal{H}_{\textup{SRG}}(\mathcal{S}_2)$, i.e., \eqref{eq: feedback_uniform_gain} does not hold. 
\end{proof}

We illustrate Theorem~\ref{th:general} through a nonlinear example.
\begin{example}
\normalfont Let $\mathcal{S}_1 = \mathbb{D}$ and $\mathcal{S}_2 = \overline{\mathbb{D}}$. Let $\varphi:\mathbb{R}\to\mathbb{R}$, $\varphi(x):=x/({1+|x|})$, and for $k \in \mathbb{R}$ let
$M_k$ denote the static operator $(M_ku)(t)=-k\varphi(u(t))$, $t\geq0$. Consider the class 
\begin{equation*}
  \mathcal{M}=\{M_k\mid k\in[0,1]\}.
\end{equation*}
    It follows directly that $\textup{SRG}(\mathcal{M}) \subseteq \mathbb{D}$. However, there is no $\epsilon>0$ such that
  $-H_1\in\mathcal{H}_{\textup{SRG}}(\mathcal{S}_1^\epsilon)$. Indeed, for $H_1 = M_1$, the inputs
  $$
  u_1(t)=0\quad \textup{and}\quad u_2(t) = \begin{cases}
      a, \quad \textup{if } t\in[0,T],\\
      0, \quad \textup{otherwise},
  \end{cases}
 $$
 with  $a \in \mathbb{R}\setminus\left\{0\right\}$ and $T>0$ generate the SRG points
  $$
 \left\{z, \bar{z}\right\} = \left\{\left(\frac{1}{1+|a|}\right)e^{\pm j\arccos\left(\frac{1}{\sqrt{1+|a|}}\right)}\right\},$$
  which tend to $1$ as $a\to0$. The set $\mathcal{S}_{2,\delta}$ is $\textup{SRG}(\mathcal{M})$-realizable for arbitrary $\delta >0$. Take $H_2 = 1+\delta$, with $\delta>0$. Then, there exist $u_1,u_2\in \mathcal{L}_2$, such that $u_1-u_2 - (1+\delta)(\varphi(u_1)-\varphi(u_2)) = 0$. Set, e.g., the inputs $u_1$ and $u_2$ as before, with $a = \delta$. Since $\mathcal{S}_2$ is star-shaped with respect to $B(0,1)$, it follows that $[H_1,H_2]$ is not uniformly incrementally bounded for all $H_2 \in \mathcal{H}_{\textup{SRG}}(\mathcal{S}_2)$. \hfill$\Box$
\end{example}

The operator-based SRG-realization property in Theorem~\ref{th:general} may impose strong structural constraints on the sets $\mathcal{S}_1$ and $\mathcal{S}_2$, making it difficult to verify in general. By exploiting additional structure, however, converse conditions can become more tractable. In the next sections, we specialize to disjunctive and conjunctive system classes.

\subsection{{Disjunctive Case}}
Consider the SRG-based system class $\mathcal{H}_{\textup{SRG}}({\mathcal{S}}^\cup_{\mathbf{\Pi}})$ in \eqref{eq:HS}, where the disjunctive set ${\mathcal{S}}^\cup_{\mathbf{\Pi}}$ in \eqref{eq:SPId} is generated by the family of matrices 
\begin{equation}\label{eq:PIPI}
    \mathbf{\Pi}=\left\{\begin{bmatrix}
        r^2-c^2 &c \\
        c & -1
    \end{bmatrix}, \:c \in \Lambda \subseteq \mathbb{R},\: r> 0\right\}.
\end{equation}
Each matrix $\Pi \in \mathbf{\Pi}$ in \eqref{eq:PIPI} characterizes the set
\begin{equation}
 \hspace{-2mm}  \left\{z \in \mathbb{C} \;\middle|\; \begin{bmatrix}
            1\\z
        \end{bmatrix}^*\Pi \begin{bmatrix}
            1\\z
        \end{bmatrix}\geq 0 \right\} = \left\{z \in \mathbb{C} \;\middle|\; |z-c|\leq r\right\},
\end{equation}
which corresponds to a disk in the complex plane centered at $c+j0$ with radius $r$. 

The disjunctive set $\mathcal{S}_{\mathbf{\Pi}}^\cup$ is star-shaped with respect to a ball, if $r^2-c^2\geq 0$ for all $\Pi \in \mathbf{\Pi}$, and $r^2-c^2>0$ for at least one $\Pi \in \mathbf{\Pi}$. We next establish a relevant realization property of the disjunctive set $\mathcal{S}_{\mathbf{\Pi}}^\cup$.

\begin{lemma}\label{lem:int}
    Let $\mathcal{L}$ be the set of causal and bounded LTI systems. Let $\mathbf{\Pi}$ be defined in \eqref{eq:PIPI}. Then, the disjunctive set $\mathcal{S}_{\mathbf{\Pi}}^\cup$ is $\textup{SRG}(\mathcal{L})$-realizable.
 \end{lemma}

\begin{proof}
   To establish SRG-realizability per Definition~\ref{def:rel_gen}, let $(Mx,x) \in \mathcal{L}_2 \times \mathcal{L}_2$ with $M \in \mathcal{L}$ be an admissible SRG point satisfying $(Mx,x)\mapsto \hat{z}\in\mathcal{S}_{\mathbf{\Pi}}^\cup$.

Recall that the set $\mathcal{S}_{\mathbf{\Pi}}^\cup$ is composed of the union of disks centered at $(c+j0)$ with radius $r$. Let the point $(Mx,x)\mapsto \hat{z}$ be located within such a disk. Note that this implies
\begin{equation}\label{eq:xy0}
 \begin{split}
&\int_{0}^\infty\begin{bmatrix}
    Mx\\x
\end{bmatrix}^\top\left(\Pi\otimes I\right)\begin{bmatrix}
    Mx\\x
\end{bmatrix}dt\\
&\qquad\qquad=r^2\|Mx\|^2-\|x-cMx\|^2\geq 0.
\end{split}
\end{equation}
Since $M \in \mathcal{L}$, by \cite[Proposition 4]{Jonsson01}, \eqref{eq:xy0} is equivalent to
\begin{equation}\label{eq:xyz}
 \begin{split}
&\int_{-\infty}^\infty\begin{bmatrix}
   M(j\omega) \hat{x}(j\omega)\\\hat{x}(j\omega)
\end{bmatrix}^*\left(\Pi\otimes I\right)\begin{bmatrix}
   M(j\omega)\hat{x}(j\omega)\\\hat{x}(j\omega)
\end{bmatrix}d\omega\geq 0.
\end{split}
\end{equation}
Since the integral is non-negative, there must exist $\omega_0 \in \mathbb{R}$ such that 
 \begin{equation}\label{eq:xy}
 \begin{split}
&\tilde{x}^*\begin{bmatrix}
   M(j\omega_0) \\I
\end{bmatrix}^*\left(\Pi\otimes I\right)\begin{bmatrix}
     M(j\omega_0)\\I
\end{bmatrix}\tilde{x}\geq 0,
\end{split}
\end{equation}
with $\tilde{x} = \hat{x}(j\omega_0)$. Define a signal $u\in \mathcal{L}_2$ having its Fourier coefficients $\hat{u}(j\omega)$ chosen such that
\begin{equation}
    \hat{u}(j\omega) = \begin{cases}
        \tfrac{1}{\sqrt{2\epsilon_0}}\tilde{x} & \textup{if } |\omega + \omega_0| < \epsilon_0, \\
         \tfrac{1}{\sqrt{2\epsilon_0}}\tilde{x}^* & \textup{if } |\omega - \omega_0| < \epsilon_0,\\
        0 &\textup{otherwise}.
    \end{cases}
\end{equation}
Then, for $\epsilon_0 \to 0$, $y = Mu$ satisfies
\begin{equation}\label{eq:xyz}
 \begin{split}
&\int_{-\infty}^\infty\begin{bmatrix}
   \hat{y}(j\omega)\\\hat{u}(j\omega)
\end{bmatrix}^*\left(\Pi\otimes I\right)\begin{bmatrix}
  \hat{y}(j\omega)\\\hat{u}(j\omega)
\end{bmatrix}d\omega\\
&=\tfrac{1}{2\epsilon_0}\int_{\omega_0-\epsilon_0}^{\omega_0+\epsilon_0}\tilde{x}^*\begin{bmatrix}
   M(j\omega)\\I
\end{bmatrix}^*\left(\Pi\otimes I\right)\begin{bmatrix}
  M(j\omega)\\I
\end{bmatrix}\tilde{x}d\omega.
\end{split}
\end{equation}
By continuity of $M(j\omega)$ at $\omega_0$, we find
\begin{equation}\label{eq:v01}
 \begin{split}
   & \lim_{\epsilon_0\to 0}\tfrac{1}{2\epsilon_0} \int_{\omega_0-\epsilon_0}^{\omega_0+\epsilon_0}\tilde{x}^*\begin{bmatrix}
   M(j\omega)\\I
\end{bmatrix}^*\left(\Pi\otimes I\right)\begin{bmatrix}
  M(j\omega)\\I
\end{bmatrix}\tilde{x}d\omega\\
&\qquad\quad= \tilde{x}^*\begin{bmatrix}
   M(j\omega_0)\\I
\end{bmatrix}^*\left(\Pi\otimes I\right)\begin{bmatrix}
 M(j\omega_0)\\I
\end{bmatrix}\tilde{x} \geq 0.
\end{split}
\end{equation}
Hence, the input-output pair $(Mu,u)\mapsto z \in \mathcal{S}_{\mathbf{\Pi}}^\cup$.

Suppose without loss of generality that $\|\hat{y}(j\omega_0)\| =\|M(j\omega_0)\hat{u}(j\omega_0)\|= 1$. Denote the complex vectors $w = \hat{y}(j\omega_0)$ and $v = \hat{u}(j\omega_0)-c\hat{y}(j\omega_0)$. Write $v$ and $w$ in the form
\begin{equation}
\begin{split}
   v^* &= \begin{bmatrix}v_{11}e^{j\theta_1}& v_{12} e^{j\theta_2}&\cdots& v_{1n}e^{j\theta_n}\end{bmatrix},\\
    w^* &= \begin{bmatrix}
    w_{11}e^{j\phi_1}&w_{12}e^{j\phi}&\cdots&w_{1n}e^{j\phi_n}
    \end{bmatrix},
    \end{split}
\end{equation}
where $v_{1i}\in \mathbb{R}$ and $w_{1j}\in \mathbb{R}$ are chosen so that $\theta_i,\phi_j \in [-\pi,0)$ for all $i,j$. Choose $\beta_i\geq 0$ and $\alpha_i \geq 0$ such that
\begin{equation}
    \angle\left(\frac{\beta_i-j\omega_0}{\beta_i+j\omega_0} \right)=\theta_i, \quad \angle\left(\frac{\alpha_j-j\omega_0}{\alpha_j+j\omega_0} \right)=\phi_j
\end{equation}
for $i=1,2,\ldots,n$ and $j=1,2,\ldots, n$. Now let
\begin{equation}
    H(s) = cI+{V(s)W(s)}\in \mathcal{RH}_{\infty}^{n\times n},
\end{equation}
where   
\begin{equation*}
    V(s) = \begin{bmatrix}v_{11}\tfrac{\alpha_1-s}{\alpha_1+s}\\\dots\\ v_{1n}\tfrac{\alpha_n-s}{\alpha_n+s}\end{bmatrix},
   W(s) = \begin{bmatrix}w_{11}\tfrac{\beta_1-s}{\beta_1+s}&\dots& w_{1n}\tfrac{\beta_n-s}{\beta_n+s}\end{bmatrix}.
\end{equation*}
Note that with this construction, we find $H(j\omega_0)\hat{y}(j\omega_0)=\hat{u}(j\omega_0)$ as $W(j\omega_0)\hat{y}(j\omega_0) = W(j\omega_0)w(j\omega_0) = \|\hat{y}(j\omega_0)\|=1$. Therefore, as $\epsilon_0\to 0$ we have $Hy = HMu = u$. 

Since $\|W(j\omega)\|=\|y\|$ and $\|V(j\omega)\|=\|v\|=\|u-cy\|$, by \eqref{eq:v01} we have $\|VW\| =\|V\|\|W\|=\|u-cy\|\leq r$, and thus $\textup{SRG}(VW)$ lies in the disk centered at the origin with radius $r$. It follows from the SRG composition rules (see, e.g., \cite[Proposition 5]{Chaffey23}) that $\textup{SRG}(H) = \textup{SRG}(cI+VW) = c+\textup{SRG}(VW)$ and thus $\textup{SRG}(H) \subseteq \left\{z\in \mathbb{C}\mid |z-c|\leq r\right\}$. Hence, $H \in \mathcal{H}_{\textup{SRG}}(\mathcal{S}_{\mathbf{\Pi}}^\cup)$, which completes the proof. 
\end{proof}

The proof of Lemma~\ref{lem:int} is constructive, and shows that the corresponding realization is an LTI system.

Associated with the disjunctive set in \eqref{eq:PIPI}, we define the conjunctive set
$\mathcal{S}_{\mathbf{\Theta}}^\cap=
(\mathbb{C}\setminus\mathcal{S}_{\mathbf{\Pi}}^\cup)^{-1}$ generated by the family
\begin{equation}\label{eq:TH}
\mathbf{\Theta}
=
\left\{
-L^\top\Pi L
\;\middle|\;
\Pi\in\mathbf{\Pi}
\right\},
\qquad
L=
\begin{bmatrix}
0&I\\
I&0
\end{bmatrix}.
\end{equation}
By construction, the corresponding regions satisfy
\begin{equation}\label{sets}
\begin{split}
\mathcal{S}_{\mathbf{\Theta}}^\cap
&=
\bigcap_{\Theta\in\mathbf{\Theta}}
\left\{
z\in\mathbb{C}
\;\middle|\;
\begin{bmatrix}1\\z\end{bmatrix}^{*}
\Theta
\begin{bmatrix}1\\z\end{bmatrix}
>0
\right\}
\\
&=
\bigcap_{\Pi\in\mathbf{\Pi}}
\left\{
z\in\mathbb{C}
\;\middle|\;
\begin{bmatrix}z\\1\end{bmatrix}^{*}
\Pi
\begin{bmatrix}z\\1\end{bmatrix}
<0
\right\}.
\end{split}
\end{equation}
We show next a property of the conjunctive set $\mathcal{S}_{\mathbf{\Theta}}^\cap$ for LTI systems, which allows to connect the operator-based SRG class with its frequency-wise counterpart in \eqref{eq:FWSRG}. This will be useful later for simplifying converse conditions for linear-nonlinear interconnections.

\begin{lemma}\label{lem:th4}
Let $\mathcal{L}$ be the set of causal and bounded LTI systems. Let $\mathbf{\Theta}$ be defined in \eqref{eq:TH}. Then: $$\mathcal{H}_{\textup{SRG}}(\mathcal{S}^\cap_{\mathbf{\Theta}})\cap \mathcal{L} = \mathcal{H}_{\textup{SRG}}^\omega(\mathcal{S}^\cap_{\mathbf{\Theta}}).$$
\end{lemma}

\begin{proof}
Let $H \in \mathcal{L}$. Since $\textup{SRG}_\omega(H) \subseteq \textup{SRG}(H)$, we find that any $H \in \mathcal{H}_{\textup{SRG}}(\mathcal{S}^\cap_{\mathbf{\Theta}})$ must also belong to $\mathcal{H}_{\textup{SRG}}^\omega(\mathcal{S}^\cap_{\mathbf{\Theta}})$. Therefore, $H \in \mathcal{H}_{\textup{SRG}}(\mathcal{S}^\cap_{\mathbf{\Theta}}) \implies H\in \mathcal{H}_{\textup{SRG}}^\omega(\mathcal{S}^\cap_{\mathbf{\Theta}})$.

We proceed by contradiction. Consider $H$ satisfying $H \in \mathcal{H}_{\textup{SRG}}^\omega(\mathcal{S}^\cap_{\mathbf{\Theta}})$, and suppose $H \not\in \mathcal{H}_{\textup{SRG}}(\mathcal{S}^\cap_{\mathbf{\Theta}})$. From the supposition, it follows that there exists a point $z_0 \in \textup{SRG}(H)$ that falls outside $\mathcal{S}^\cap_{\mathbf{\Theta}}$. Since $\mathcal{S}_{\mathbf{\Theta}}^\cap
=
(\mathbb{C}\setminus\mathcal{S}_{\mathbf{\Pi}}^\cup)^{-1}$, the fact that $z_0\notin\mathcal{S}_{\mathbf{\Theta}}^\cap$ implies that for some $\Pi\in\mathbf{\Pi}$,
$$
\begin{bmatrix}z_0\\1\end{bmatrix}^*
\Pi
\begin{bmatrix}z_0\\1\end{bmatrix}
\geq0,
$$
and thus $r|z_0|\geq |cz_0-1|$. From the definition of $\textup{SRG}(H)$, and in particular $z_0(u,y) = \rho(u,y)e^{\pm j \theta(u,y)}$ with $\rho$ and $\theta$ in \eqref{eq:gain} and \eqref{eq:phase} respectively, we find that
\begin{equation*}
\int_{-\infty}^\infty\!\!\big(\|rH_1(j\omega)\hat{u}(j\omega)\|^2 - \|(cH_1(j\omega)+I)\hat{u}(j\omega)\|^2\big)d\omega \geq 0,
\end{equation*}
where we have used $\|u\|=1$. This, in turn, implies the existence of a $\omega_0 \in \mathbb{R}$ and $x \in \mathbb{C}^n$ such that
\begin{equation}\label{eq:vw}
x^*\begin{bmatrix}
    H(j\omega_0)\\I
\end{bmatrix}^*\left(\begin{bmatrix}
    r^2-c^2 & c\\c &-1
\end{bmatrix}\otimes I\right)\begin{bmatrix}
    H(j\omega_0)\\I
\end{bmatrix}x\geq 0.
\end{equation}
Hence $(x,H(j\omega_0)x)\mapsto z$ satisfies $z \not \in \mathcal{S}_{\mathbf{\Theta}}^\cap$, and thus $H \not \in \mathcal{H}_{\textup{SRG}}^\omega(\mathcal{S}_{\mathbf{\Theta}}^\cap)$. The supposition can therefore not hold true, and thus $H \in \mathcal{H}_{\textup{SRG}}^\omega(\mathcal{S}^\cap_{\mathbf{\Theta}}) \implies H\in \mathcal{H}_{\textup{SRG}}(\mathcal{S}^\cap_{\mathbf{\Theta}})$.

Combining the above two implications, we find that
$$H \in \mathcal{H}_{\textup{SRG}}(\mathcal{S}^\cap_{\mathbf{\Theta}}) \iff H \in \mathcal{H}_{\textup{SRG}}^\omega(\mathcal{S}^\cap_{\mathbf{\Theta}}),$$
and thus $\mathcal{H}_{\textup{SRG}}(\mathcal{S}^\cap_{\mathbf{\Theta}}) = \mathcal{H}_{\textup{SRG}}^\omega(\mathcal{S}^\cap_{\mathbf{\Theta}})$.
\end{proof}

Equipped with the above properties, we formulate an \emph{operator-based} converse result for linear-nonlinear feedback interconnections, where $H_1$ is LTI and $H_2$ is a general system (linear or nonlinear). 

\begin{theorem}\label{th:converse220}
 Let $\mathbf{\Pi}$ be defined in \eqref{eq:PIPI} and $\mathbf{\Theta}$ be the associated family in \eqref{eq:TH}. Suppose the following holds:
\begin{enumerate}
    \item $H_1\in \mathcal{L}$;
    \item $r^2-c^2\geq 0$, not all zero, for all $\Pi \in \mathbf{\Pi}$;
    \item $[H_1, H_2]$ is well-posed for all $H_2 \in \mathcal{H}_{\textup{SRG}}(\mathcal{S}^\cup_{\mathbf{\Pi}})$. 
\end{enumerate}
Then, $[H_1, H_2]$ is uniformly incrementally bounded for all $H_2 \in \mathcal{H}_{\textup{SRG}}(\mathcal{S}^\cup_{\mathbf{\Pi}})$ if and only if there exists $\epsilon>0$ such that $-H_1 \in \mathcal{H}_{\textup{SRG}}((\mathcal{S}^\cap_{\mathbf{\Theta}})^\epsilon)$. 
\end{theorem}

\begin{proof}
We show that all conditions of Theorem~\ref{th:general} are satisfied. First, note that $\big[\begin{smallmatrix}
0\\1\end{smallmatrix}\big]^\top\Pi\big[\begin{smallmatrix}
        0\\1
    \end{smallmatrix}\big]=-1\leq 0$ for all $\Pi \in \mathbf{\Pi}$ with $\mathbf{\Pi}$ in \eqref{eq:PIPI} ensures $0 \in \mathcal{S}^\cap_\mathbf{\Theta}$. Next, let $\mathcal{M} = \mathcal{L}$. Condition $1)$ in Theorem~\ref{th:general} is satisfied by supposition 1). From supposition 2) it follows that $H_2 \in \mathcal{H}_{\textup{SRG}}(\mathcal{S}^\cup_{\mathbf{\Pi}})$ implies $\tau H_2 \in \mathcal{H}_{\textup{SRG}}(\mathcal{S}^\cup_{\mathbf{\Pi}})$ for all $\tau \in (0,1]$ and thus condition 2) is satisfied. Combining this with supposition 3) shows that the system $[H_1, \tau H_2]$ is well-posed for all $\tau\in (0,1]$, and condition $5)$ holds. Moreover, by supposition 2) the set $\mathcal{S}_{\mathbf{\Pi}}^\cup$ is star-shaped with respect to a ball, and so condition 3) is satisfied. Since $\mathcal{S}_{\mathbf{\Pi,\delta}}^\cup$ is a disjunctive set that results from inflation, it follows directly from Lemma~\ref{lem:int} that $\mathcal{S}_{\mathbf{\Pi,\delta}}^\cup$ is $\textup{SRG}(\mathcal{L})$-realizable, hence condition 5) is satisfied. 
\end{proof}

We highlight a few aspects of Theorem~\ref{th:converse220} and its proof.
\begin{enumerate}
\item The proof of Theorem~\ref{th:converse220} relies on constructing a destabilizing system $H_2$ within the class $\mathcal{H}_{\textup{SRG}}(\mathcal{S}_{\mathbf{\Pi}}^\cup)$. Although this class contains both linear and nonlinear operators, it follows from the constructive proof of $\textup{SRG}(\mathcal{L})$-realizability (Lemma~\ref{lem:int}) that the destabilizing system can be selected as an LTI system.  

     \item When $H_1$ is a single-input single-output system, Lemma~\ref{lem:th4} provides a simple graphical interpretation of Theorem~\ref{th:converse220}: the Nyquist plot $\left\{-H_1(j\omega)\mid \omega\in\mathbb{R}\right\}$ must remain within the region $\mathcal{S}_{\mathbf{\Theta}}^\cap$. For $|\mathbf{\Pi}|=1$, this is akin to the classical circle criterion \cite[Chapter V]{Desoer09}. That is, when $r^2-c^2>0$ the Nyquist plot should remain inside of a disk with radius $r'=r/(r^2-c^2)$ and center $c'+j0$, where $c'=-c/(r^2-c^2)$, and for $r^2-c^2=0$ it should remain to the right of the point $-1/(2c)$. Different from the circle criterion, however, we are not limited to sector-bounded nonlinearties.
    \item An analogue of the general result in Theorem~\ref{th:general} can be formulated in terms of IQCs by replacing the SRG classes with corresponding IQC classes. As a consequence of the previously established SRG-IQC class equivalences in Lemma~\ref{lem:eq} for disjunctive and conjunctive sets, that is, $\mathcal{H}_{\textup{SRG}}(\mathcal{S}^\cup_{\mathbf{\Pi}})=\mathcal{H}^\cup_{\textup{IQC}}(\mathbf{M}_{\mathbf{\Pi}})$ and $\mathcal{H}_{\textup{SRG}}(\mathcal{S}^\cap_\mathbf{\Theta})=\mathcal{H}^\cap_{\textup{IQC}}(\mathbf{M}_{\mathbf{\Theta}})$, it directly follows that Theorem~\ref{th:converse220} admits an \emph{equivalent} formulation in terms of IQCs.  
\end{enumerate}

From Theorem~\ref{th:converse220} we recover special SRG versions of the converse small-gain and passivity theorems, which mirror those of \cite{Khong18} under a single constraint.

\begin{corollary}[Converse passivity]
    Suppose $H_1\in \mathcal{L}$ and let $\mu >0$. Then, $[H_1, H_2]$ is incrementally bounded for all $H_2\in \mathcal{B}_\Delta$ satisfying $\textup{SRG}(H_2) \subseteq \left\{z \in \mathbb{C} \mid \textup{Re}\left\{z\right\} \geq \mu |z|^2\right\}$, if and only if there exists $\epsilon >0$ such that $\textup{SRG}(H_1) \subseteq \left\{z \in \mathbb{C} \mid \textup{Re}\left\{z\right\} \geq -\mu+\epsilon\right\}$. 
\end{corollary}

\begin{proof}
    This is Theorem~\ref{th:converse220} with $\mathbf{\Pi}=\left\{\begin{bmatrix}
        0 & \tfrac{1}{2\mu}\\ \tfrac{1}{2\mu} &-1
    \end{bmatrix}\right\}$.
\end{proof}

\begin{corollary}[Converse small-gain]
    Suppose $H_1\in \mathcal{L}$ and let $\alpha>0$. Then, $[H_1, H_2]$ is incrementally bounded for all $H_2\in\mathcal{B}_\Delta$ satisfying $\textup{SRG}(H_2) \subseteq \left\{z \in \mathbb{C}\mid |z|\leq \alpha\right\}$, if and only if there exists $\epsilon >0$ such that $\textup{SRG}(H_1) \subseteq \left\{z \in \mathbb{C}\mid |z|\leq \tfrac{1}{\alpha}-\epsilon\right\}$. 
\end{corollary}

\begin{proof}
    This is Theorem~\ref{th:converse220} with $\mathbf{\Pi}=\left\{\begin{bmatrix}
        \alpha^2  & 0\\ 0 &-1
    \end{bmatrix}\right\}$.
\end{proof}

For multiple constraints, we obtain a converse mixed small-gain/passivity theorem for LTI systems, which parallels the results in \cite{Liu25}.

\begin{corollary}[Converse mixed small-gain/passivity]
    Suppose $H_1\in \mathcal{L}$ and let $\mu>0,\, \alpha>0$. Then, $[H_1,H_2]$ is incrementally bounded for all $H_2\in \mathcal{B}_\Delta$ satisfying
    \begin{equation*}
        \textup{SRG}(H_2) \subseteq \left\{z \in \mathbb{C} \mid \textup{Re}\left\{z\right\}\geq \epsilon |z|^2\right\} \cup \left\{z \in \mathbb{C} \mid |z|\leq \alpha \right\},
    \end{equation*}
    if and only if there exists $\epsilon >0$ such that
     \begin{equation*}
     \begin{split}
       & \textup{SRG}(H_1) \subseteq\\
       &\quad\left\{z \in \mathbb{C} \mid \textup{Re}\left\{z\right\}\geq -\mu+\epsilon\right\} \cap \left\{z \in \mathbb{C} \mid |z|\leq \tfrac{1}{\alpha}-\epsilon \right\}.
        \end{split}
    \end{equation*}
\end{corollary}

\begin{proof}
    This is Theorem~\ref{th:converse220} with
    $$\mathbf{\Pi} = \left\{\begin{bmatrix}
            0 & \tfrac{1}{2\mu}\\\tfrac{1}{2\mu}&-1
        \end{bmatrix}, \begin{bmatrix}
            \alpha^2 & 0\\ 0 &-1
        \end{bmatrix}\right\}. 
        $$
\end{proof}

We present an application of Theorem~\ref{th:converse220} within a robust control design example.

\begin{example}
     \normalfont Consider an uncertain plant $P = P_0+\Delta$, where $P_0$ captures the nominal LTI dynamics given by
     $$
     P_0(s) = \frac{1}{s^2+2s+1},
     $$
    and where $\Delta$ is a (possibly nonlinear) uncertain component that is assumed to belong to $\mathcal{H}_{\textup{SRG}}(\mathcal{S}_{\mathbf{\Pi}}^\cup)$, with 
     $$
     \mathbf{\Pi}= \left\{\begin{bmatrix}
            0 & \tfrac{1}{2}\\\tfrac{1}{2}&-0.2
        \end{bmatrix}, \begin{bmatrix}
            1 & 0\\ 0 &-1
        \end{bmatrix}\right\}, 
     $$
     i.e., $\Delta$ introduces additional passive and/or finite-gain dynamics, see Fig.~\ref{fig:Nyq} (left). The plant is interconnected with an LTI feedback controller $C$ according to $y = Pu$, and $u = C(w-y)$, where $w \in \mathcal{L}_2$ is an exogenous input. The resulting closed-loop system can be reformulated to fit the context of Fig.~\ref{fig:FB1}, with 
     $$
     H_1(s) = \frac{C(s)}{1+P_0(s)C(s)}, \:\:\textup{ and }\:\: H_2 = \Delta.
     $$
     To guarantee robust stability with respect to all $\Delta \in \mathcal{H}_{\textup{SRG}}(\mathcal{S}_{\mathbf{\Pi}}^\cup)$, it follows from Theorem~\ref{th:converse220} that $C$ must be designed such that $H_1$ is stable, and belongs to the region $\mathcal{S}_{\mathbf{\Theta}}^\cap$ as depicted in Fig.~\ref{fig:Nyq} (right). A possible design that realizes this is given by the lead filter
     $$
     C_{\textup{lead}}(s)= \frac{1}{s+1}+0.12.
     $$
     \begin{figure}[htbp]
    \centering
\includegraphics[width=.45\textwidth]{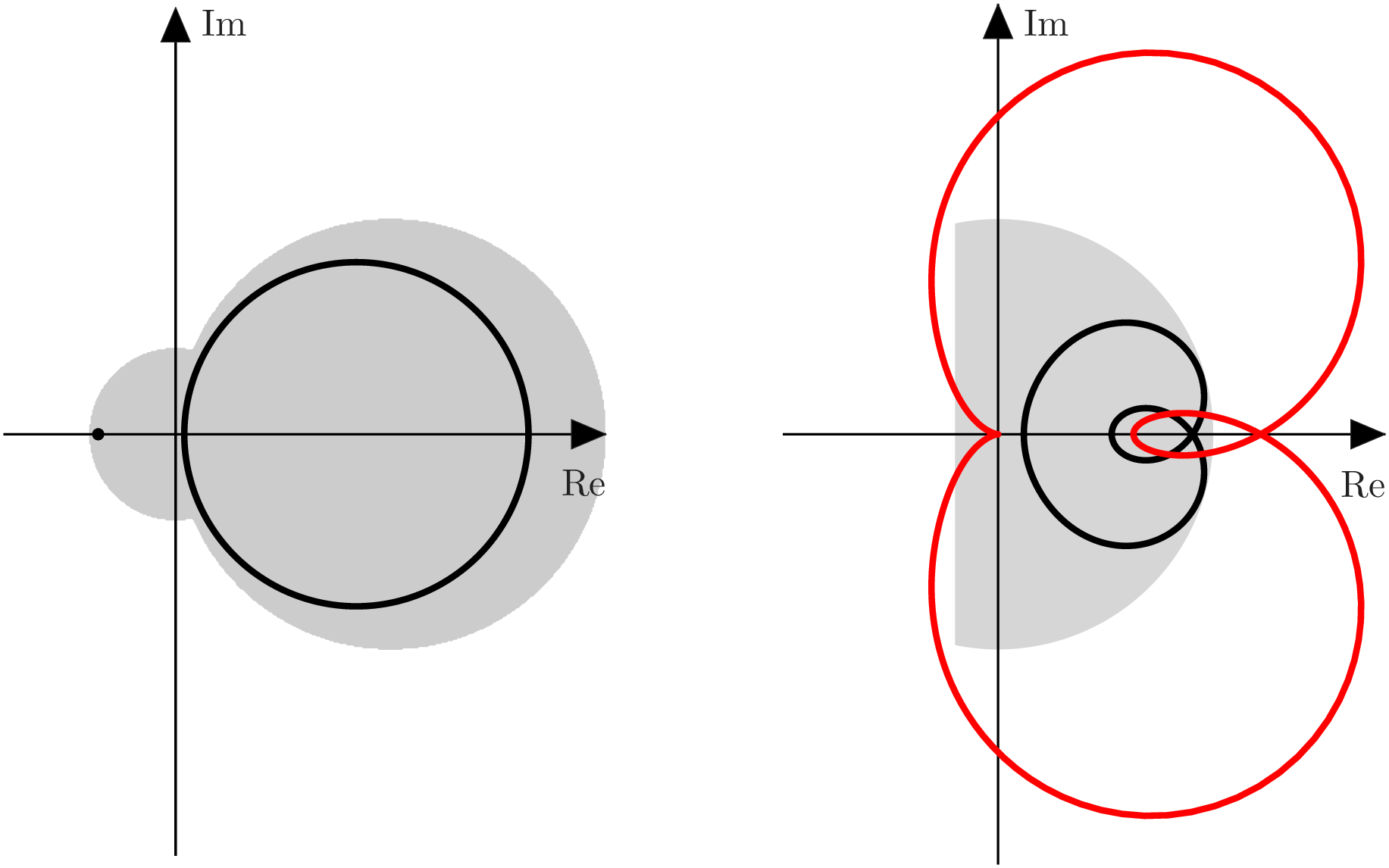}
    \caption{Left: uncertainty region $\mathcal{S}_{\mathbf{\Pi}}^\cup$ (grey) and Nyquist diagrams of destabilizing uncertainties $\Delta$ (black). Right: plant region $\mathcal{S}_{\mathbf{\Theta}}^\cap$ (grey) and Nyquist diagrams of $H_1$ with $C_{\textup{lead}}$ (black) and with $C_{\textup{lag}}$ (red).}
    \label{fig:Nyq}
\end{figure}
     The resulting Nyquist diagram of $H_1$ is shown in Fig.~\ref{fig:Nyq} (in black), and indeed shows that this system is passive and has small-gain. As an alternative, consider the lag filter
     $$
     C_{\textup{lag}}(s) = \frac{2}{s^2+2s+1}.
     $$
     Although the resulting system $H_1$ is stable, it is not passive, nor does it have a small gain, as can be seen from the Nyquist diagram in Fig.~\ref{fig:Nyq} (in red). As a consequence, the closed-loop system is not robustly stable to $\Delta$. Destabilizing uncertainties for this case can be constructed as
     \begin{equation}
         \Delta(s) =\frac{4}{s+1}+0.1,\: \textup{ and }\: \Delta(s) = -0.9,
     \end{equation}
     which both belong to $\mathcal{H}_{\textup{SRG}}(\mathcal{S}_{\mathbf{\Pi}}^\cup)$, see also Fig.~\ref{fig:Nyq}.
     \hfill$\Box$
\end{example}

We close the disjunctive case by specializing to the full LTI setting, yielding the following frequency-wise converse result.

\begin{theorem}\label{th:converse0}
 Let $\mathbf{\Pi}$ be defined in \eqref{eq:PIPI} and $\mathbf{\Theta}$ be the associated family in \eqref{eq:TH}. Suppose the following holds:
\begin{enumerate}
    \item $H_1, H_2\in \mathcal{L}$;
    \item $r^2-c^2\geq 0$ for all $\Pi \in \mathbf{\Pi}$;
    \item $[H_1, H_2]$ is well-posed for all $H_2 \in \mathcal{H}_{\textup{SRG}}^\omega(\mathcal{S}^\cup_{\mathbf{\Pi}})$. 
\end{enumerate}
Then, $[H_1, H_2]$ is incrementally bounded for all $H_2 \in \mathcal{H}_{\textup{SRG}}^\omega(\mathcal{S}^\cup_{\mathbf{\Pi}})$ if and only if $-H_1 \in \mathcal{H}_{\textup{SRG}}^\omega(\mathcal{S}^\cap_{\mathbf{\Theta}})$. 
\end{theorem}

\begin{proof}
The proof follows the same steps as the proof of Theorem~\ref{th:converse220}, combined with Lemma~\ref{lem:th4} and the construction of a destabilizing LTI system $H_2$ as in the proof of Lemma~\ref{lem:int}. 
\end{proof}

The set of admissible LTI systems $H_2$ in Theorem~\ref{th:converse220} is a subset of that in Theorem~\ref{th:converse0}, while the corresponding sets of systems $H_1$ are identical. Hence, Theorem~\ref{th:converse0} establishes robustness with respect to a \emph{larger} LTI uncertainty class, owing to the structured frequency-wise characterization. 

\subsection{Conjunctive Case}
A direct counterpart of Theorem~\ref{th:converse220} for the case where $-H_1 \in \mathcal{H}_{\textup{SRG}}(\mathcal{S}_{\mathbf{\Theta}}^\cup)$ and $H_2 \in \mathcal{H}_{\textup{SRG}}(\mathcal{S}^\cap_\mathbf{\Pi})$ is generally \emph{false}. The prime reason is that the conjunctive set $\mathcal{S}_{\mathbf{\Pi}}^\cap$ is not necessarily SRG($\mathcal{L}$)-realizable, and establishing realizability in general remains an open problem (note that $\mathcal{S}_{\mathbf{\Pi}}^\cap$ can be star-shaped). In the full LTI setting, however, this obstruction can be alleviated by introducing a frequency-wise notion of SRG-realizability.

\begin{definition}[Frequency-wise SRG-realizability]\label{def:rel_gen}
  Let $\mathcal{S}\subseteq\mathbb{C}$ and $\mathcal{M}\subseteq\mathbb{C}^{n\times n}$.
We say that $\mathcal{S}$ is \emph{$\mathrm{SRG}_\omega(\mathcal{M})$-realizable} if, for every $M\in\mathcal{M}$ and some $0\neq u\in\mathbb{C}^n$, such that $(Mu,u)\mapsto z \in \mathcal{S}$, there exists $H(s) \in \mathcal{RH}_{\infty}^{n\times n}$ satisfying $u = H(j\omega_0)Mu$ for some $\omega_0 \in \mathbb{R}$ and $\textup{SRG}_\omega(H) \subseteq \mathcal{S}$. When $\mathcal{M}=\mathbb{C}^{n\times n}$ we simply say that $\mathcal{S}$ is \emph{$\textup{SRG}_\omega$-realizable}. 
\end{definition}

Frequency-wise SRG-realizability means that every SRG point contained in $\mathcal{S}$ that is realized as the frequency-response point of an LTI system, admits a causal and bounded LTI realization whose entire frequency-wise SRG remains in $\mathcal{S}$. Adopting this notion, we obtain the following frequency-wise converse result for LTI systems.

\begin{theorem}\label{th:converse02}
 Let $\mathbf{\Pi}$ be defined in \eqref{eq:PIPI} and let $\mathbf{\Theta}$ be the associated family in \eqref{eq:TH}. Let $\delta_0>0$. Suppose the following holds:
\begin{enumerate}
    \item $H_1, H_2\in \mathcal{L}$, and $H_1(s) \in \mathcal{M}$;
    \item $r^2-c^2\geq 0$, not all zero, for all $\Pi \in \mathbf{\Pi}$;
    \item $(\mathcal{S}^\cap_{\mathbf{\Pi}})_\delta$ is $\textup{SRG}_\omega(\mathcal{M})$-realizable for all $\delta \in (0,\delta_0]$; 
    \item $[H_1, H_2]$ is well-posed for all $H_2 \in \mathcal{H}^\omega_{\textup{SRG}}(\mathcal{S}^\cap_{\mathbf{\Pi}})$. 
\end{enumerate}
Then, $[H_1, H_2]$ is incrementally bounded for all $H_2 \in \mathcal{H}^\omega_{\textup{SRG}}(\mathcal{S}^\cap_{\mathbf{\Pi}})$ if and only if there exists $\epsilon >0$ such that $-H_1 \in \mathcal{H}^\omega_{\textup{SRG}}((\mathcal{S}^\cup_{\mathbf{\Theta}})^\epsilon)$. 
\end{theorem}

\begin{proof}
    The proof of \emph{sufficiency} follows the same steps as in the proof of Theorem~\ref{th:general}, now invoking the frequency-wise SRG result in Theorem~\ref{th:fw}.

For the proof of \emph{necessity}, we proceed in the same way as in the proof of Theorem~\ref{th:general}, using now frequency-wise SRG-realizability as per Definition~\ref{def:rel_gen}, as well as star-shapedness of $\mathcal{S}_{\mathbf{\Pi}}^\cap$, which follows from the fact that by supposition 2) in the theorem, $\mathcal{S}_{\mathbf{\Pi}}^\cap$ is a convex set.
\end{proof}

A difficulty in Theorem~\ref{th:converse02} remains in verifying frequency-wise $\textup{SRG}$-realizability of $\mathcal{S}^\cap_{\mathbf{\Pi}}$. The case $|\mathbf{\Pi}|=1$ follows immediately from Lemma~\ref{lem:int} as in that case $\mathcal{S}^\cap_{\mathbf{\Pi}} = \mathcal{S}^\cup_{\mathbf{\Pi}}$; hence, the nontrivial challenge arises when $|\mathbf{\Pi}|>1$. To gain more insights into the possible structure of frequency-wise $\textup{SRG}$-realizable conjunctive sets, we provide an explicit example.

\begin{example}\label{ex:1}
  \normalfont  Consider the transfer function
    \begin{equation}
        F(s) = \frac{1}{s^3+3s^2+2s+1}-0.1736.
    \end{equation}
    Let $\omega^* \in \mathbb{R}_{\geq 0}$ be such that $\textup{Re}(F(j\omega^*)) = 0$ and $\textup{Im}(F(j\omega^*)) =\max_{\omega \in \mathbb{R}_{\geq 0}}\textup{Im}(F(j\omega))$. Define the function 
    $$
r(\lambda) :=
\max_{\omega \in [0,\omega^*]}|F(j\omega)+|\lambda||, \quad \lambda \in \mathbb{R},
$$
and the parametrized matrices
$$
\Pi_\lambda :=
\begin{bmatrix}
r(\lambda)^2-\lambda^2 & \lambda\\
\lambda & -1
\end{bmatrix}, \quad \lambda \in \mathbb{R}.
$$
We will show that the associated conic region
$$
\mathcal{S}_{\mathbf{\Pi}}^\cap
:=
\bigcap_{\lambda \in \mathbb{R}}
\left\{
z \in \mathbb{C} \;\middle|\;
\begin{bmatrix}1 \\ z\end{bmatrix}^*
\Pi_\lambda
\begin{bmatrix}1 \\ z\end{bmatrix}
\ge 0
\right\}
$$
is $\textup{SRG}_\omega(\mathcal{M})$-realizable with $\mathcal{M}=\left\{dI \mid d \in \mathbb{C}\right\}$, i.e., diagonal systems. An illustration of the set $\mathcal{S}_{\mathbf{\Pi}}^\cap$ is given in Fig.~\ref{fig:sets2} along with $F(j\omega)$ (in black).

\begin{figure}[htbp]
    \centering
\includegraphics[ width=.35\textwidth]{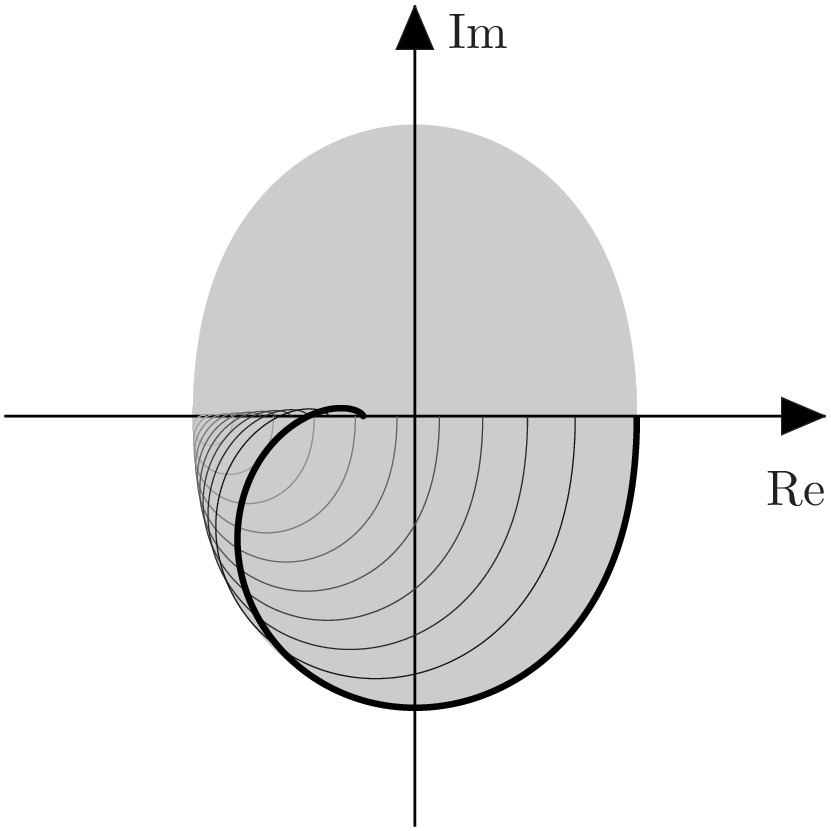}
    \caption{Conic set $\mathcal{S}_{\mathbf{\Pi}}^\cap$ with $F(j\omega)$ (in black). The different realizations $H(j\omega) = kF(j\omega)+c$ are shown in dark grey.}
    \label{fig:sets2}
\end{figure}
It can be readily verified that the boundary of $\mathcal{S}_{\mathbf{\Pi}}^\cap$ is given by $\partial \mathcal{S}_{\mathbf{\Pi}}^\cap = \Gamma \cup (-\Gamma) \cup \Gamma^* \cup (-\Gamma^*)$,
with 
$$
\Gamma = \left\{F(j\omega) \mid \omega \in [0,\omega^*]\right\}.
$$
The choices $H(s) = F(s)$ and $H(s)=-F(s)$ realizes systems whose frequency-responses cover all points on the arcs $\Gamma$ and $-\Gamma$, and for which it clearly follows from Fig.~\ref{fig:sets2} that $\textup{SRG}_{\omega}(H) \subseteq \mathcal{S}_{\mathbf{\Pi}}^\cap$. It remains to show that each point on $\Gamma^*$ and $-\Gamma^*$ can be realized by a causal, stable system $H$ that satisfies $\textup{SRG}_{\omega}(H) \subseteq \mathcal{S}_{\mathbf{\Pi}}^\cap$. For this purpose, we consider the system $H(s) = kF(s)+c$, with $k,c\in \mathbb{R}$. For each scaling $k \in [0,1]$, we can find the minimum horizontal distance $c$ between $kF(j\omega)$ and $-\Gamma^*$, so that $H(j\omega)$ yields a point on $-\Gamma^*$. All points on $\Gamma^*$ are realised with $-H(s)$. Note that by construction, $\textup{SRG}(H) \subseteq \mathcal{S}_{\mathbf{\Pi}}^\cap$. All interior points in $\mathcal{S}_{\mathbf{\Pi}}^\cap$ can be reached by scaling $H \to \tau H$ with $\tau \in [0,1]$.

Since the SRG of $H(s)I$ is identical to that of $H(s)$, we find that $\mathcal{S}_{\mathbf{\Pi}}^\cap$ is $\textup{SRG}(\mathcal{M})$-realizable. \hfill$\Box$
\end{example}

The observations in Example~\ref{ex:1} motivate the conjecture that sufficiently smooth sets whose boundaries are generated by functions $F(s) \in \mathcal{RH}_{\infty}^{n\times n}$ are $\textup{SRG}_{\omega}$-realizable. We leave the rigorous analysis of this conjecture for future work.

\section{Conclusions}\label{sec:conclusions}
In this paper we have established equivalences between the IQC and SRG frameworks, and developed converse results. Key in our results is the introduction of novel disjunctive and conjunctive system classes, characterized by unions and intersections of quadratic input–output constraints. Our main findings show that, under mild assumptions, the corresponding graph separation conditions in the IQC and SRG frameworks are equivalent. By exploiting this equivalence together with the disjunctive and conjunctive system classes, we derived new converse theorems for linear time-invariant and nonlinear systems described by multiple constraints, thereby extending the scope of non-conservative robustness analysis. The converse theory is complete for the disjunctive case. The conjunctive case presents fundamental structural challenges that remain unresolved and motivate future research.

\bibliographystyle{IEEEtran}

\end{document}